\documentclass[aps,superscriptaddress,amsmath,amsfonts,showpacs]{revtex4}

\usepackage{graphicx}
\usepackage{bm}

\begin{document}

\title{
Topological Phases of Noncentrosymmetric Superconductors:
Edge States, Majorana Fermions, and
the Non-Abelian Statistics
}


\author{Masatoshi Sato}
\affiliation{The institute for Solid State Physics, The University of
Tokyo, Kashiwanoha 5-1-5, Kashiwa-shi, Chiba 277-8581, Japan}
\author{Satoshi Fujimoto}
\affiliation{Department of Physics, Kyoto University, Kyoto 606-8502, Japan}


%\email[fuji@scphys.kyoto-u.ac.jp] 
%\thanks{}


\date{\today}

\begin{abstract}
The existence of edge states and zero energy modes in vortex cores 
is a hallmark of topologically
nontrivial phases realized in various
condensed matter systems such as the fractional quantum Hall states,
$p+ip$ superconductors, and ${\bm Z}_2$ insulators (quantum spin Hall state).
We examine this scenario
for two dimensional noncentrosymmetric superconductors 
which allow the parity-mixing of
Cooper pairs.
It is found that even when the $s$-wave pairing gap is nonzero, 
provided that the superconducting gap of 
spin-triplet pairs is larger than that of spin-singlet pairs,
gapless edge states and zero energy Majorana modes in vortex cores
emerge, characterizing topological order.
Furthermore, it is shown that for Rashba superconductors,
the quantum spin Hall effect produced by gapless edge states exists  
even under an applied magnetic field which breaks time-reversal symmetry
provided that the field direction is perpendicular to
the propagating direction of the edge modes.
This result making a sharp contrast to the ${\bm Z}_2$ insulator
is due to an accidental symmetry inherent in the Rashba model.
It is also demonstrated that in the case with magnetic fields,
the non-Abelian statistics of vortices
is possible under a particular but realistic condition.
\end{abstract}

%\pacs{74.90.+n, 74.25.Ha, 73.43.-f, 71.10.Pm}


\maketitle

%\section{}
%\subsection{}
%\subsubsection{}

\section{Introduction}

Topological phases of matter are quantum many-body states
with topologically non-trivial structures of the Hilbert spaces \cite{wen}.
They are characterized by the existence of 
both topologically-protected gapless modes on the edges and
 a bulk energy gap which
separates the ground state 
and excited states \cite{wen2,moore,nayak,fradkin,hatsugai}.
Recently, topological phases are of interest
in connection with wide ranges of subjects in condensed matter physics, such as
the quantum Hall effect \cite{TKNN},
$p+ip$ superconductors \cite{moore,nayak,fradkin,read,ivanov,stern}, and
${\bm Z}_2$ topological insulators 
(quantum spin Hall effect) \cite{KM05a,SSTH05,BZ06}.
The gapless edge states are topologically stable against the perturbations
that do not break symmetries of systems, and play
a crucial role in transport properties such as 
quantum (spin) Hall effects, which is, currently, inspiring
applications to spintronics \cite{wen2,BZ06,zhang2}.
The total number of topologically-protected edge modes in a given system is
associated with topological numbers, i.e. 
the TKNN number (the first Chern number of the U(1) bundle describing
the many-body wave function) for systems without time-reversal 
symmetry \cite{TKNN,hatsugai}, and the ${\bm Z}_2$ invariant 
(the parity of the spin-resolved TKNN number)
for time-reversal invariant systems \cite{KM05a}.

The presence of low-energy edge modes also implies
the fractionalization of quasiparticles in the bulk \cite{lee}.
For instance, in a vortex core of a spinless $p+ip$ superconductor,
there is a zero energy mode which is described by a Majorana fermion, i.e.
a half of a conventional fermion \cite{kopnin}.
A vortex with a Majorana fermion is a quasiparticle
obeying the non-Abelian statistics \cite{read,ivanov,Sato03,stone1,stern}:
The braiding of vortices with Majorana fermions gives rise to
the superposition of the degenerate many-body ground states.
Owing to this entangled character, non-Abelian anyons can be
utilized for the construction of fault-tolerant
quantum computers \cite{freedman,stone1,das,tewari2,das2}.
When there are odd number of vortices with Majorana fermions,
the edge state must be also Majorana, since an isolated Majorana fermion
is unphysical, and should be paired with another Majorana fermion \cite{stone2}.
In this sense, the Majorana edge state is a concomitant of
a zero energy Majorana mode in a vortex core, and
the existence of both of them characterizes the topological order.

A more precise argument is given in terms of the low-energy effective theory.
The low-energy effective theory for topological phases is 
the Chern-Simons theory, 
and both edge states and
topological quasiparticles in the bulk are described by 
the corresponding conformal field theory. 
In the case with the level-$k$ SU(2) symmetry, it is
the SU(2)$_k$ Wess-Zumino-Witten theory, which is decomposed into
the U(1) gaussian theory with the central charge $c=1$ 
and the ${\bm Z}_k$ parafermion theory with
$c=\frac{2(k-1)}{k+2}$ \cite{moore,nayak,fradkin}.
For superconductors, where quasiparticles are superpositions of particles and
holes, the U(1) gaussian part describing fractional charges for
the fractional quantum Hall state is absent, and thus,
the low-energy effective theory for edge states and quasiparticles
is the ${\bm Z}_k$ parafermion theory.
The topological phase of $p+ip$ superconductors corresponds to
$k=2$, i.e. the Ising conformal field theory, the operator content of which
includes Majorana fermions \cite{stone1,kitaev}. 

The search for possible realization of topological phases
is an intriguing and challenging issue involving the development of
novel concepts for 
quantum condensed states as well as potential technological applications.
In this paper, we demonstrate that two dimensional (2D) noncentrosymmetric
superconductors (NCS) under certain circumstances provide
another candidate for the physical realization of
a topological phase, which
supports the existence of edge states and zero energy modes in vortex cores.
In NCS, asymmetric spin-orbit (SO) interactions play crucial roles in
various exotic superconducting phenomena such as
parity-mixing of Cooper pairs, magnetoelectric effects, 
and helical vortex phases \cite{ede,ede2,gr,yip2,fri,sam2,kau,fuji3,fuji4}. 
It is found that the asymmetric SO interaction is also a key
to realize topological phases in NCS.
We investigate the edge states and the zero energy vortex core states
in NCS by using both numerical and analytical methods, and
verify the condition for the realization of topological phases in NCS.

The main results of this paper are as follows.
For 2D NCS with the admixture of $s$-wave pairing
and $p$-wave pairing, as long as the $p$-wave gap is larger than
the $s$-wave gap, topological phases are realized.
The classification of the topological phases in NCS is completed by 
examining topological numbers.
In the absence of magnetic fields,
a ${\bm Z}_2$ topological phase with two gapless edge modes emerges, 
which leads to the quantum spin Hall effect;
i.e. when an electric field or a temperature gradient is applied,
a spin Hall current carried by the edge state flows in the
direction perpendicular to the applied external field. 
Moreover, it is shown that in the case of the Rashba SO interaction,
the gapless edge modes are stable against weak magnetic field 
applied perpendicular
to the direction in which the edge modes propagate.
This is a bit surprising because the magnetic field which breaks 
time-reversal symmetry flaws
the ${\bm Z}_2$ classification of the topological phase, and also
the TKNN number is still zero for such a weak magnetic field.
In fact, the stability of this topological phase is due to an accidental
symmetry for particular symmetry points in the Brillouin Zone specific to
the Rashba model, which is characterized by another topological number,
 ``winding number''.
We will clarify the condition for the nonzero winding number in this paper.
In addition to these spin Hall states,
there is also a topological phase with the nonzero TKNN number,
which is analogous to the quantum Hall state, and realized for a particular
electron density with a magnetic field.
The implication of these gapless edge states for experimental
observations will be also discussed.

Furthermore, we examine the vortex core state for these topological
phases by solving the Bogoliubov-de Gennes equations.
On the basis of both numerical and analytical methods,
it is found that
zero energy Majorana fermion modes exist in the vortex cores.
It is also verified how the non-Abelian statistics of vortices 
is realized in NCS.
The non-Abelian statistics of vortices in superconductors has been
considered so far for the chiral $p+ip$ state.\cite{read,ivanov}
In the case of spinless $p+ip$ state, there is only one zero energy
Majorana mode in a vortex, which is a non-Abelian anyon.
In the case of spinful $p+ip$ state, 
there are two Majorana fermions in a vortex corresponding to spin up and
spin down states, which form a usual complex fermion. The non-Abelian
statistics is not realized for this situation.
As Ivanov elucidated,
when there is a half quantum vortex which suppresses one of two Majorana fermions
in a vortex core, vortices behave as non-Abelian anyons \cite{ivanov}.
For NCS, however, this scenario is not applicable, because 
the half-quantum vortex is a texture of the $\mbox{\boldmath $d$}$-vector
of $p$-wave pairing, and the $\mbox{\boldmath $d$}$-vector in NCS
is locked by the strong SO interaction.
As proposed by one of the authors \cite{fuji1},
when one tunes a chemical potential
so as that the Fermi level crosses the $\Gamma$ point in
the Brillouin zone at which the electron band is a Dirac cone, and
applies
a Zeeman field parallel to the $z$-direction, 
a Majorana fermion associated with the Dirac cone is eliminated, and
only one Majorana fermion survives in a vortex
core, which makes the non-Abelian statistics of vortices possible.
We examine this strategy by using an analytical approach based on
the index theorem for zero modes developed by Tewari, Sarma and Lee
in the case of $p+ip$ superconductors \cite{tewari}.
It is found that when the spin-triplet gap is larger than the spin-singlet gap,
the non-Abelian statistics of vortices is realized 
for the situation mentioned above.

It would be useful to comment on some recent other studies related to
the present paper.
1) Some topological properties of the
Rashba type NCS, such as the nodal structure of the gap function and the
edge state, were discussed in \cite{Sato06,TYBN08}.
While these literatures assumed the time-reversal symmetry and
were based on the ${\bm Z}_2$ topological number,
we complete the topological classification by taking into account the
TKNN number simultaneously.
We also find that the ``winding number'' mentioned above
is useful to characterize topological phases of the Rashba superconductors.
2) In \cite{fuji1}, 
a possibility of zero energy states in vortex cores
for purely $s$-wave noncentrosymmetric superconductors 
under strong magnetic fields was suggested.
Later, unfortunately, it was turned out that the zero energy vortex core 
state found in \cite{fuji1} is spurious, and there is no
zero energy state in purely s-wave cases.
The misleading conclusion in \cite{fuji1} is due to the erroneous
assumption that the unitary transformation which diagonalizes the spin index
is commutative with the center of mass coordinate of Cooper pairs raised
by the presence of vortices. This assumption is valid only when
the quasiclassical approximation is justified.
This is not the case for the issue of zero energy vortex core states.
This point will be discussed in more detail in Sec. \ref{subsec:chiralitybasis}
of the current paper.
3) Independently, Lu and Yip elucidated that in NCS with parity-mixing,
the zero energy vortex core state is possible only when the gap
of the spin-triplet pairs is larger than that of the spin-singlet
pairs \cite{yip}. 
They examined the condition for zero energy modes of the Bogoliubov-de Gennes
(BdG) equations without obtaining explicit solutions.
However, to investigate the possible realization of the non-Abelian statistics
of vortices, it is desirable to derive the explicit zero energy solutions
which enable us to see whether a zero energy mode in a vortex core is 
a Majorana fermion or a usual complex fermion.
This issue is addressed in the current paper.

The organization of this paper is as follows.
In Sec. \ref{sec:topology}, we start with a general topological
argument, which enables us to understand edge and vortex core states clearly,
providing the complete classification of topological phases in NCS.  
In Sec. \ref{sec:edge}, gapless edge states in NCS are investigated by using 
the general topological argument and numerical methods.
The implication for transport properties associated with edge states
is also discussed.
In Sec. \ref{sec:vortex}, we explore zero energy vortex core states 
for NCS on the basis of numerical solutions 
for the Bogoliubov-de Gennes equations and 
an analytical approach based on the index theorem for zero energy states, 
and clarify the condition for the realization of the non-Abelian
statistics of vortices.
We conclude in Sec. \ref{sec:summary} with a summary of our results.

\section{Topological numbers for noncentrosymmetric superconductors}
\label{sec:topology}

In this paper, we consider type II NCS with Rashba-type SO interaction \cite{rash}
in two dimensions.
For concreteness, we define our model on the square lattice, though
the following argument does not rely on the particular choice of the
crystal structure.
Then the model Hamiltonian is 
\begin{eqnarray}
{\mathcal H}&=&\sum_{{\bm k},\sigma}\varepsilon_{\bm k}
 c_{{\bm k}\sigma}^{\dagger}c_{{\bm k}\sigma}
-\mu_{\rm B}H_z \sum_{{\bm k},\sigma}(\sigma_z)_{\sigma\sigma'}
 c_{{\bm k}\sigma}^{\dagger}c_{{\bm k}\sigma}
+\alpha\sum_{{\bm k},\sigma,\sigma'}{\bm {\mathcal L}}_0({\bm k})
\cdot{\bm \sigma}_{\sigma\sigma'}c_{{\bm k}\sigma}^{\dagger}c_{{\bm k}\sigma'} 
\nonumber\\
&+&\frac{1}{2}\sum_{{\bm k}\sigma\sigma'}
\Delta_{\sigma\sigma'}({\bm k})c_{{\bm k}\sigma}^{\dagger}
c_{-{\bm k}\sigma'}^{\dagger}
+\frac{1}{2}\sum_{{\bm k}\sigma\sigma'}
\Delta_{\sigma'\sigma}^*({\bm k})c_{-{\bm k}\sigma}c_{{\bm k}\sigma'}, 
\label{eq:Hamiltonian}
\end{eqnarray}
where $c_{{\bm k}\sigma}^{\dagger}$ ($c_{{\bm k}\sigma}$) is a creation (an
annihilation) operator for an electron with momentum ${\bm
k}=(k_x,k_y)$, spin $\sigma$.
The energy band dispersion is $\varepsilon_{\bm k}=-2t(\cos k_x+\cos
k_y)-\mu$ with the hopping parameter $t$ and the chemical potential
$\mu$, and the Rashba SO coupling is ${\bm {\mathcal L}}_0({\bm k})=(\sin k_y, -\sin k_x)$.
For simplicity, we assume that $t>0$ and $\mu<0$.
Because of parity mixing of Cooper pairs, 
the gap function $\Delta({\bm k})$ has both a spin-triplet component
${\bm d}({\bm k})$ and a spin-singlet one $\psi({\bm k})$ at the same time,
$\Delta({\bm k})=i\psi({\bm k})\sigma_y+i{\bm d}({\bm k}){\bm \sigma}\sigma_y$. 
Due to the strong SO coupling, the spin-triplet component ${\bm d}({\bm k})$ is
aligned with the Rashba coupling, ${\bm d}({\bm k})=\Delta_{\rm t}{\bm {\mathcal
L}_0}({\bm k})$ \cite{fri}.
For the spin-singlet component $\psi$, we assume a $s$-wave pairing,
$\psi({\bm k})=\Delta_{\rm s}$. 
The amplitudes $\Delta_{\rm t,s}$ are chosen as real.
The Zeeman coupling
$\mu_{\rm B}H_z\sum_{\bm k}
(c^{\dagger}_{{\bm k}\uparrow}c_{{\bm k}\uparrow}
-c^{\dagger}_{{\bm k}\downarrow}c_{{\bm k}\downarrow})$ 
with $H_z$ a magnetic field in the $z$ direction
has been also 
introduced for later use.

Before going to study topological properties of the system, we first
examine the bulk spectrum of the system. 
Topological nature of the system changes only when
the gap of the bulk spectrum closes.
The bulk spectrum $E({\bm k})$ of the system is obtained by diagonalizing the
following matrix, 
\begin{eqnarray}
H({\bm k})=
 \left(
\begin{array}{cc}
\varepsilon_{\bm k}-\mu_{\rm B}H_z\sigma_z+\alpha{\bm {\mathcal
 L}}_0({\bm k})\cdot
{\bm \sigma} &
i\Delta_{\rm s}\sigma_y+i\Delta_{\rm t}{\bm {\mathcal L}}_0({\bm k})
\cdot{\bm \sigma}\sigma_y\\
-i\Delta_{\rm s}\sigma_y
-i\Delta_{\rm t}{\bm {\mathcal L}}_0({\bm k})\sigma_y\cdot{\bm \sigma}&
- \varepsilon_{\bm k}+\mu_{\rm B}H_z\sigma_z+\alpha{\bm {\mathcal
L}}_0({\bm k})
\cdot{\bm \sigma}^{*}
\end{array}
 \right),
\end{eqnarray}
and we have
\begin{eqnarray}
E({\bm k})=\pm\sqrt{\varepsilon_{\bm k}^2+(\alpha^2+\Delta_{\rm t}^2){\bm
{\mathcal L}}_0({\bm k})^2+\mu_{\rm B}^2H_z^2+\Delta_{\rm s}^2
\pm2\sqrt{
(\varepsilon_{\bm k}\alpha
+\Delta_{\rm s}\Delta_{\rm t})^2{\bm {\mathcal L}}_0({\bm k})^2
+(\varepsilon_{\bm k}^2+\Delta_{\rm s}^2)\mu_{\rm B}^2H_z^2}
}.
\end{eqnarray}
The gap of the system closes only when 
\begin{eqnarray}
\varepsilon_{\bm k}^2+(\alpha^2
+\Delta_{\rm t}^2){\bm {\mathcal L}}_0({\bm k})^2
+\mu_{\rm B}^2H_z^2+\Delta_{\rm s}^2
=2\sqrt{(\varepsilon_{\bm k}\alpha
+\Delta_{\rm s}\Delta_{\rm t})^2{\bm {\mathcal L}}_0({\bm k})^2
 +(\varepsilon_{\bm k}^2+\Delta_{\rm s}^2)\mu_{\rm B}^2H_z^2}, 
\end{eqnarray}
which is equivalent to
\begin{eqnarray}
\varepsilon_{\bm k}^2+\Delta_{\rm s}^2
=\mu_{\rm B}^2H_z^2+(\alpha^2+\Delta_{\rm t}^2){\bm {\mathcal L}}_0({\bm
k})^2,
\quad
\varepsilon_{\bm k}\Delta_{\rm t}{\bm {\mathcal L}}_0({\bm k})
=\Delta_{\rm s}\alpha{\bm {\mathcal L}}_0({\bm k}).
\label{eq:gap-closing}
\end{eqnarray}
When $\Delta_{\rm t}\neq 0$, (\ref{eq:gap-closing}) is met either
when  
\begin{eqnarray}
\varepsilon_{\bm k}=\frac{\Delta_{\rm s}}{\Delta_{\rm t}}\alpha,
\quad 
\left(1+\frac{\alpha^2}{\Delta_{\rm t}^2}\right)
\left(\Delta_{\rm t}^2{\bm {\mathcal L}}_0({\bm k})^2-\Delta_{\rm
 s}^2\right)+\mu_{\rm B}^2H_z^2=0.
\label{eq:gap-closing1-1}
\end{eqnarray}
or
\begin{eqnarray}
\varepsilon_{\bm k}^2+\Delta_s^2=\mu_{\rm B}^2H_z^2,
\quad
{\bm {\mathcal L}}_0({\bm k})=0.
\label{eq:gap-closing1-2}
\end{eqnarray}
In the absence of the magnetic field, only Eqs.(\ref{eq:gap-closing1-1}) can be
met and they are rewritten in simpler forms,
\begin{eqnarray}
\varepsilon_{\bm k}^2=\alpha^2{\bm {\mathcal L}}_0({\bm k})^2,
\quad
\Delta_{\rm t}^2{\bm {\mathcal L}}_0({\bm k})^2=\Delta_{\rm s}^2.
\label{eq:gap-closing2}
\end{eqnarray}
Topological nature of the system does not change unless
(\ref{eq:gap-closing1-1}) or (\ref{eq:gap-closing1-2}) (or
(\ref{eq:gap-closing2}) when $H_z=0$) is satisfied.

\subsection{${\bm Z}_2$ topological number}
\label{subsec:z2}

When $H_z=0$, the system is time-reversal invariant, and the
topological property is characterized by the ${\bm Z}_2$ invariant
\cite{KM05a, KM05b, MB07, FKM07, Roy06, FK07}, \footnote{In this paper,
we concentrate on topological properties in two dimensions.
For three dimensional time-reversal invariant
superconductors, there exists another topological number \cite{SRFL08}}.
We will show that if the spin-triplet pairing is stronger than the spin-singlet
one, the ${\bm Z}_2$ number is non-trivial.

To calculate the ${\bm Z}_2$ number, we adiabatically deform the
Hamiltonian of the system without gap closing. 
This process does not change the ${\bm Z}_2$ topological number, 
since it changes only when the gap closes.
From (\ref{eq:gap-closing2}), it is found that if
the spin-triplet amplitude $\Delta_{\rm t}{\bm {\mathcal L}}_0({\bm k})$
is larger than the spin-singlet
one $\Delta_{\rm s}$ on the Fermi surface given by
$\varepsilon_{\bm k}=\alpha{\bm {\mathcal L}}_0({\bm k})$, we can 
take $\Delta_{\rm s}\rightarrow 0$, then $\alpha\rightarrow 0$ without gap
closing.
(If $\varepsilon_{\bm k}=0$ at one of the time-reversal momenta ${\bm k}=(0,0),
(\pi,0), (0,\pi), (\pi,\pi)$, the gap closes when $\Delta_{\rm
s}=0$. However, this undesirable gap closing can be avoided by
changing $\mu$ or $t$ slightly.)
Thus, its ${\bm Z}_2$ number is the same as that of the
pure spin-triplet SC with ${\bm d}({\bm k})=\Delta_{\rm t}{\bm {\mathcal
L}}_0({\bm k})$.

As was shown in \cite{Sato08},
the ${\bm Z}_2$ topological number $(-1)^{\nu}$ for a pure spin-triplet SC
is determined by its dispersion $\varepsilon_{\bm k}$ in the normal state,
\begin{eqnarray}
(-1)^{\nu}=\prod_{{\bm k}=(0,0),(0,\pi),(\pi,0),(\pi,\pi)}
 {\rm sgn}\varepsilon_{\bm k}. 
\end{eqnarray}
From this formula, it is easily shown that the ${\bm Z}_2$
number is always non-zero (mod 2) for the square lattice system.  
Therefore, from the argument above, if the spin-triplet pairing is stronger
than the singlet one, the NCS is a topological insulator with a
non-trivial ${\bm Z}_2$ number.
This topological superconductor belongs to the same class as one
discussed in \cite{QHRZ08,Roy,TYBN08}.

\subsection{TKNN number}
In the presence of a magnetic field, 
the time-reversal invariance is broken.
The ${\bm Z}_2$ topological number is no longer meaningful, and the TKNN
number plays a central role in topological nature of the system instead.
In the following argument, we consider only the Zeeman effect of the
magnetic field,
neglecting the orbital depairing effect.
When the energy scale of the SO interaction is sufficiently larger than the Zeeman energy scale,
the Pauli depairing effect does not exit for magnetic
fields parallel to the $z$-axis, and the structure of the ${\bm d}$-vector for
the spin-triplet pairs is not affected by the Zeeman effect \cite{fri,fuji4}.
We assume such situations.
The effect of the orbital depairing effect will be discussed at the end of
this subsection.

In order to obtain a non-zero TKNN number, the magnetic field
should be large enough to have a gap closing. Otherwise the TKNN number
must be zero since the system is adiabatically connected to the case
with $H_z=0$.
 (When $H_z=0$, the TKNN number becomes zero because of time-reversal
invariance.)
When the spin-triplet pairing is stronger than the singlet one,
(\ref{eq:gap-closing1-1}) is never met in the presence of the magnetic
field, and a gap closing occurs only when (\ref{eq:gap-closing1-2}) is
satisfied.
Since ${\bm {\mathcal L}}_0({\bm k})$ is zero at ${\bm k}=(0,0),
(0,\pi), (\pi,0),
(\pi,\pi)$, it is found that we have a gap closing if one of the following
equations is satisfied,
\begin{eqnarray}
(-4t-\mu)^2+\Delta_{\rm s}^2=\mu_{\rm B}^2H_{z}^2,
\quad
\mu^2+\Delta_{\rm s}^2=\mu_{\rm B}^2H_{z}^2,
\quad
(4t-\mu)^2+\Delta_{\rm s}^2=\mu_{\rm B}^2H_{z}^2.
\end{eqnarray}

\begin{figure}[h]
\begin{center}
\includegraphics[width=12cm]{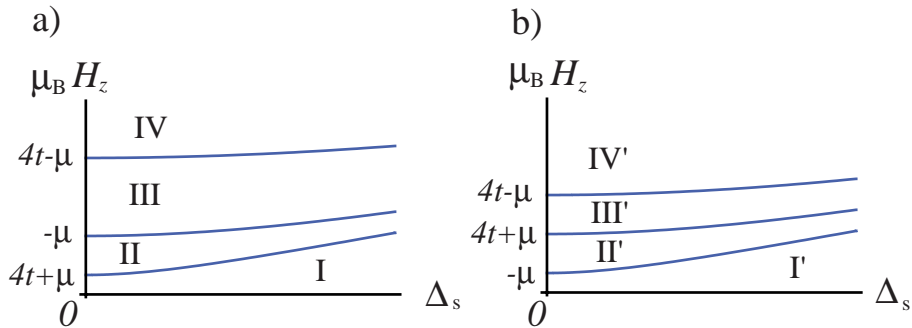}
\caption{Schematic picture of phase diagram of 2D NCS with Rashba coupling.
 }
\label{fig:phase}
\end{center}
\end{figure}

From this, it is found that we have a two classes of phase diagram
illustrated in Fig.\ref{fig:phase}.
a) For $-4t <\mu<-2t$, the first phase transition occurs at
$\mu_{\rm B}H_z=\sqrt{(4t+\mu)^2+\Delta_{\rm s}^2}$, and the second one
at $\mu_{\rm B}H_z=\sqrt{\mu^2+\Delta_{\rm s}^2}$, and the third one
at $\mu_{\rm B}H_z=\sqrt{(4t-\mu)^2+\Delta_{\rm s}^2}$.
b) For $-2t <\mu< 0$, the first phase transition occurs at
$\mu_{\rm B}H_z=\sqrt{\mu^2+\Delta_{\rm s}^2}$, and the second one
at $\mu_{\rm B}H_z=\sqrt{(4t+\mu)^2+\Delta_{\rm s}^2}$, and the third one
at $\mu_{\rm B}H_z=\sqrt{(4t-\mu)^2+\Delta_{\rm s}^2}$.
The patterns of the TKNN numbers differ for these parameter regions.

The TKNN number for each phase is evaluated by using the adiabatic
deformation of the Hamiltonian in a similar manner as Sec.\ref{subsec:z2}. 
As is evident from Fig.\ref{fig:phase}, we can take $\Delta_{\rm
s}=\alpha= 0$ without crossing the phase boundary.
For $\Delta_{\rm s}=\alpha=0$,  the up-spin electron and down-spin electron
are decoupled from each other, and the Hamiltonian reduces to a pair of
$2\times 2$ matrices, 
\begin{eqnarray}
H_{\uparrow \uparrow}({\bm k})= 
\left(
\begin{array}{cc}
\varepsilon_{\bm k}-\mu_{\rm B}H_z & -i\Delta_{\rm t}(\sin k_x-i\sin k_y)
 \\
i\Delta_{\rm t}(\sin k_x+i\sin k_y) 
 & -\varepsilon_{\bm k}+\mu_{\rm B}H_z
\end{array} 
\right),
\nonumber\\
H_{\downarrow \downarrow}({\bm k})= 
\left(
\begin{array}{cc}
\varepsilon_{\bm k}+\mu_{\rm B}H_z & -i\Delta_{\rm t}(\sin k_x+i\sin k_y)
 \\
i\Delta_{\rm t}(\sin k_x-i\sin k_y) 
 & -\varepsilon_{\bm k}-\mu_{\rm B}H_z
\end{array} 
\right).
\end{eqnarray}
Using these $2\times 2$ matrices, we can evaluate the TKNN integers.
By rewriting $H_{\sigma \sigma}({\bm k})$ $(\sigma=\uparrow,
\downarrow)$ as $H_{\sigma \sigma}({\bm k})={\bm R}_{\sigma}({\bm
k})\cdot {\bm \sigma}$,
the TKNN number $I_{\rm TKNN}$ is given by
\begin{eqnarray}
I_{\rm TKNN}=-\sum_{\sigma}\frac{1}{8\pi}\int\int d^2{\bm
 k}\epsilon_{abc}\epsilon_{ij} \hat{\bm R}_{\sigma}^a \partial_i \hat{\bm R}_{\sigma}^b
 \partial_j \hat{\bm R}_{\sigma}^c, 
\end{eqnarray}  
with $\hat{\bm R}_{\sigma}={\bm R}_{\sigma}/|{\bm R}_{\sigma}|$.

In Table.\ref{table:topologicalnumber}, we summarize the obtained TKNN numbers.
The phases IV and IV' are trivial band insulators, and of no interest.
The other phases are in topologically nontrivial states.
For the phases II, III, and III', there is only one band which crosses the Fermi
level, and is associated with the nonzero TKNN number.
The sign of the TKNN number depends on the curvature of the Fermi surface
as well as the chirality of the gap function, since it 
is related to the Hall conductivity.
The gap functions for the phases II, III, and III' possess the same chirality,
and hence,
 for the positive curvature of the Fermi surface, $I_{\rm TKNN}>0$ 
(the phase II), and
for the negative curvature, $I_{\rm TKNN}<0$ (the phases III and III').
In the phase II', there are two bands with opposite signs of the curvatures
of the Fermi surfaces. Also, the gap functions for these bands have opposite
chiralities. Thus, each of the two bands 
contributes to the TKNN number equal to $-1$,
leading to the total TKNN number $I_{\rm TKNN}=-2$. 
The phases II, III, II', and III' are analogous to
the quantum Hall state.

For the phases I and I', although both the TKNN number and the ${\bm Z}_2$ index
are zero,
there exist topological orders with two gapless edge modes
and Majorana fermions in vortex cores, 
as will be shown in Sec. \ref{sec:edge} and Sec. \ref{sec:vortex},
and their topological nature is similar
to that of the ${\bm Z}_2$ insulator.
In the next subsection, it will be clarified that
these phases are characterized by another topological number, i.e. 
winding number.

In the argument above, the orbital depairing effect of magnetic fields
is ignored.
For typical superconductors, the orbital limiting field is
$H_{\rm c}^{\rm orb}\sim E_F(\Delta/E_F)^2/\mu_{\rm B}\ll t/\mu_{\rm B}$
with the Fermi energy $E_F$. 
Thus,
superconductivity
does not survive in the strong magnetic field regions, III and III', 
though the Pauli depairing effect 
is completely suppressed by the strong asymmetric SO interaction as mentioned before.
(To avoid confusion, we would like to stress again that the topological phases obtained
above by putting $\alpha=0$ can be deformed into topologically equivalent states
with nonzero large $\alpha$, unless the bulk gap closes.)
On the other hand,
the realization of the phases I, II, I', and II' in
the weak field regions is
feasible.

\begin{table}
\begin{center} 
\begin{tabular}[t]{|c|c|c|c|c|}
\multicolumn{4}{l}{a) $-4t\le \mu <-2t$} \\ 
 \hline
& $\mu_{\rm B}H_z$ &$I_{\rm TKNN}$& $I(0)$ & $I(\pi)$\\ 
\hline 
I& $0<\mu_{\rm B}H_z<\sqrt{(4t+\mu)^2+\Delta_s^2}$ &  0 &-2 &0 \\ 
II &$\sqrt{(4t+\mu)^2+\Delta_s^2}<\mu_{\rm B}H_z<\sqrt{\mu^2+\Delta_s^2}$
& 1 & -1 &0 \\
III& $\sqrt{\mu^2+\Delta_s^2}<\mu_{\rm B}H_z<\sqrt{(4t-\mu)^2+\Delta_s^2}$
& -1 & 0 &-1 \\
IV& $\sqrt{(4t-\mu)^2+\Delta_s^2} <\mu_{\rm B}H_z$ & 0& 0 & 0 \\
\hline
\end{tabular}
\begin{tabular}[t]{|c|c|c|c|c|}
\multicolumn{4}{l}{b) $-2t<\mu<0$} \\ 
 \hline
&$\mu_{\rm B}H_z$ &$I_{\rm TKNN}$& $I(0)$ & $I(\pi)$\\ 
\hline 
I'&$0<\mu_{\rm B}H_z<\sqrt{\mu^2+\Delta_s^2}$ &  0 &-2 &0 \\ 
II'&$\sqrt{\mu^2+\Delta_s^2}<\mu_{\rm B}H_z<\sqrt{(4t+\mu)^2+\Delta_s^2}$
& -2 & -1 &-1 \\
III'&$\sqrt{(4t+\mu)^2+\Delta_s^2}<
\mu_{\rm B}H_z<\sqrt{(4t-\mu)^2+\Delta_s^2}$
& -1 & 0 &-1 \\
IV'&$\sqrt{(4t-\mu)^2+\Delta_s^2}<\mu_{\rm B}H_z$ & 0& 0 & 0 \\
\hline
 \end{tabular} 
\end{center}
\caption{The TKNN integer $I_{\rm TKNN}$ and the winding number $I(k_y)$
 for 2D NCS with Rashba coupling.}
\label{table:topologicalnumber}
\end{table}

\subsection{Winding number}
\label{subsec:winding}

As was mentioned in the previous subsection, the NCS with the Rashba
coupling has an additional topological number, in addition to the ${\bm
Z}_2$ and TKNN numbers.
Unlike the usual topological number, 
the additional topological number is defined only for particular values of
momentum, but it is also useful to understand properties of edge
states and vortex core states, which are specific to the 2D NCS, in the
presence of a magnetic field.

To define the topological number,  let us consider the particle-hole
symmetry of the Hamiltonian,
\begin{eqnarray}
\Gamma H({\bm k})\Gamma^{\dagger}= -H(-{\bm k})^{*},
\quad
\Gamma =\left(
\begin{array}{cc}
0 &{\bm 1} \\
{\bm 1} & 0 
\end{array}
        \right).
\end{eqnarray}
For $k_y=0$ or $k_y=\pi$, we have $H(-{\bm k})^*=H({\bm k})$, 
thus the Hamiltonian anti-commutes with $\Gamma$,
$
\{\Gamma, H({\bm k})\}=0. 
$
This implies that if we take the basis where $\Gamma$ is diagonal
$\Gamma={\rm diag}(1,1,-1,-1)$, then 
the Hamiltonian becomes off-diagonal
\begin{eqnarray}
H({\bm k})=\left(
\begin{array}{cc}
0 & q({\bm k}) \\
q^{\dagger}({\bm k}) &0
\end{array}
\right),
\end{eqnarray}
for these values of $k_y$.
Using $q({\bm k})$, we can define the topological number as \cite{WZ89},
\begin{eqnarray}
I(k_y)=\frac{1}{4\pi i} \int_{-\pi}^{\pi}
 dk_x {\rm tr}(q^{-1}({\bm k})\partial_{k_x}q({\bm k})
 -q^{\dagger -1}({\bm k})\partial_{k_x}q^{\dagger}({\bm k})),
\quad
(k_y=0, \, \pi).
\end{eqnarray}

An explicit calculation shows that
 $q({\bm k})$ is given by
\begin{eqnarray}
q({\bm k})=
\left(
\begin{array}{cc}
\varepsilon_{\bm k}-\mu_{\rm B}H_z +i\Delta_{\rm t}\sin k_x &
 -\Delta_{\rm s}+i\alpha \sin k_x \\
\Delta_{\rm s}-i\alpha \sin k_x & \varepsilon_{\bm k}+\mu_{\rm B}H_z
 +i\Delta_{\rm t}\sin k_x
\end{array}
\right).
\end{eqnarray}
From this, we can calculate $I(k_y)$ $(k_y=0,\pi)$ for each phase in
Fig.\ref{fig:phase}.
The result is summarized in Table.\ref{table:topologicalnumber}.

The additional topological number $I(k_y)$ is accidental and
very sensitive to the direction of the magnetic field.
While $I(k_y)$ remains well-defined even in the presence of a magnetic
field in the $x$-direction,
it becomes meaningless if we apply a magnetic field in the
$y$-direction, $H_y$, since $H_y$ breaks the relation $H(-{\bm k})^*=
H({\bm k})$
for $k_y=0,\pi$.
From the bulk-edge correspondence, this means that edge states for
Rashba type NCS are also very sensitive to the direction of the magnetic
field, which will be confirmed numerically in the next section.

\section{Edge states in noncentrosymmetric superconductors}
\label{sec:edge}

In this section, we investigate edge states for the 2D NCS numerically.
From the bulk edge correspondence, a non-trivial bulk topological number
implies gapless edge states.
This will be confirmed in this section.  
Experimental implications for the gapless edge states are also discussed
in Sec.\ref{subsec:edgeexperiment}.

\subsection{Without a magnetic field}
Let us first study edge states for the 2D NCS in the absence of magnetic
field.
Consider the lattice version of the Hamiltonian (\ref{eq:Hamiltonian})
\begin{eqnarray}
{\cal H}&=&-t\sum_{\langle {\bm i},{\bm j}\rangle,\sigma}
 c_{{\bm i}\sigma}^{\dagger}c_{{\bm j}\sigma}
 -\mu \sum_{{\bm i},\sigma}c_{{\bm i}\sigma}^{\dagger}c_{{\bm i}\sigma}
-\mu_{\rm B}H_z \sum_{{\bm i},\sigma,\sigma'}(\sigma_z)_{\sigma\sigma'}
c_{{\bm i}\sigma}^{\dagger}c_{{\bm i}\sigma'}
-i\frac{\alpha}{2}
\sum_{\langle {\bm i},{\bm j}\rangle, \sigma,\sigma'}
({\bm \sigma}_{\sigma\sigma'}\times \hat{\bm r}_{{\bm i}{\bm j}})_z
c_{{\bm i}\sigma}^{\dagger}c_{{\bm j}\sigma'}
\nonumber\\
&&+\Delta_{\rm s}\sum_{\bm i}
(c_{{\bm i}\uparrow}^{\dagger}c_{{\bm i}\downarrow}^{\dagger}+{\rm h.c.})
\nonumber\\
&&-\frac{1}{2}\Delta_{\rm t}\sum_{\bm i}
(c_{{\bm i}\uparrow}^{\dagger}c_{{\bm i}+\hat{x}\uparrow}^{\dagger}+
c_{{\bm i}\downarrow}^{\dagger}c_{{\bm i}+\hat{x}\downarrow}^{\dagger}
-ic_{{\bm i}\uparrow}^{\dagger}c_{{\bm i}+\hat{y}\uparrow}^{\dagger}
+ic_{{\bm i}\downarrow}^{\dagger}c_{{\bm
i}+\hat{y}\downarrow}^{\dagger}+{\rm h.c.}),
\end{eqnarray}
where ${\bm i}=(i_x,i_y)$ denotes a site on the square lattice,
$\hat{\bm r}_{{\bm i}{\bm j}}$ a unit vector from a site ${\bm i}$
to a site ${\bm j}$. The sum $\sum_{\langle {\bm i}{\bm j}\rangle}$ is
taken between the nearest neighbor sites. 
In this subsection, we suppose $H_z=0$.
Consider the system with two edges at $i_x=0$ and $i_x=N_x$, and put 
the periodic boundary condition in the $y$ direction.
By solving numerically the energy spectrum as a function of the momentum
$k_y$ in the $y$ direction, edge states for NCS are
studied.

\begin{figure}
\includegraphics[width=11cm]{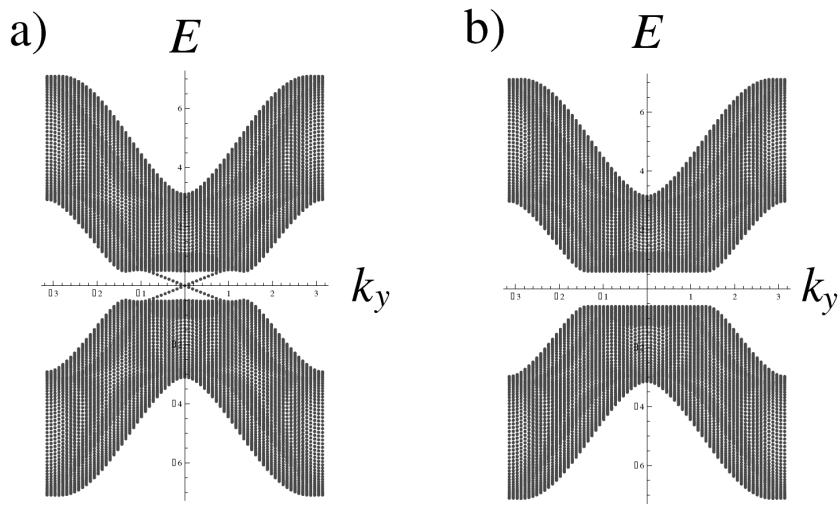}
\caption{The energy spectra of the 2D NCS with edges at $i_x=0$ and
$i_x=50$ in the absence of magnetic field.
Here $k_y$ denotes the momentum in the $y$-direction.
We take $t=1$, $\mu=-3$, $\alpha=0.6$. a) NCS with dominating $p$-wave paring.
 $\Delta_{\rm t}=0.6$ and  $\Delta_{\rm s}=0.1$. b) NCS with purely $s$-wave
 paring. $\Delta_{\rm t}=0$ and $\Delta_{\rm s}=0.6$.}
\label{fig:edgestatewithouth} 
\end{figure}

As was shown in the previous section, the ${\bm Z}_2$ topological number is 
non-zero if the gap of the spin-triplet pairs are larger than that of the singlet.  
Therefore, from the bulk-edge correspondence, 
there should always exist gapless edges if the spin-triplet pairs
dominates the superconductivity.
In Fig.\ref{fig:edgestatewithouth} a), we show the energy spectrum of
the 2D NCS with edges.
It is found that there exist gapless edge states in the bulk gap.
The gapless edges states form a Kramers pair.

For comparison, we also illustrate the energy spectrum for
the 2D NCS with purely $s$-wave paring in Fig.\ref{fig:edgestatewithouth} b).
As is seen clearly, no edge state is obtained.
This is also consistent with the trivial ${\bm Z}_2$ number of the
purely $s$-wave paring.

\subsection{With a magnetic field}
\label{subsec:withh}

Let us now examine edge states in the case with a magnetic field.
As is shown in the previous section,
there exists a variety of topological phases characterized by the
topological numbers.

\begin{figure}
\includegraphics[width=10cm]{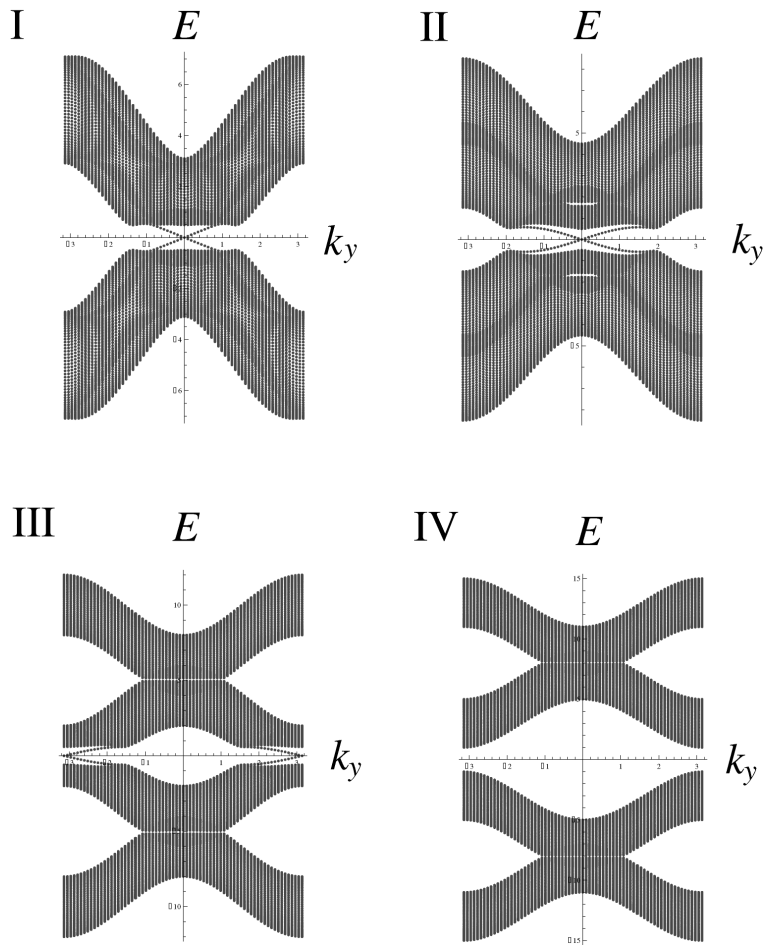}
\caption{The energy spectra of 2D NCS with edges at $i_x=0$ and
$i_x=50$ in the phases I, II, III and IV in Fig.\ref{fig:phase} a).
Here $k_y$ denotes the momentum in the $y$-direction.
We take $t=1$, $\mu=-3$, $\alpha=0.6$, $\Delta_{\rm t}=0.6$ and
 $\Delta_{\rm s}=0.1$. $H_z$ is I) $\mu_{\rm B}H_z=0$,
 II) $\mu_{\rm B}H_z=1.5$, III) $\mu_{\rm B}H_z=5$, and IV) $\mu_{\rm B}H_z=8$.}
\label{fig:edgestatehza} 
\end{figure}

\begin{figure}
\includegraphics[width=10cm]{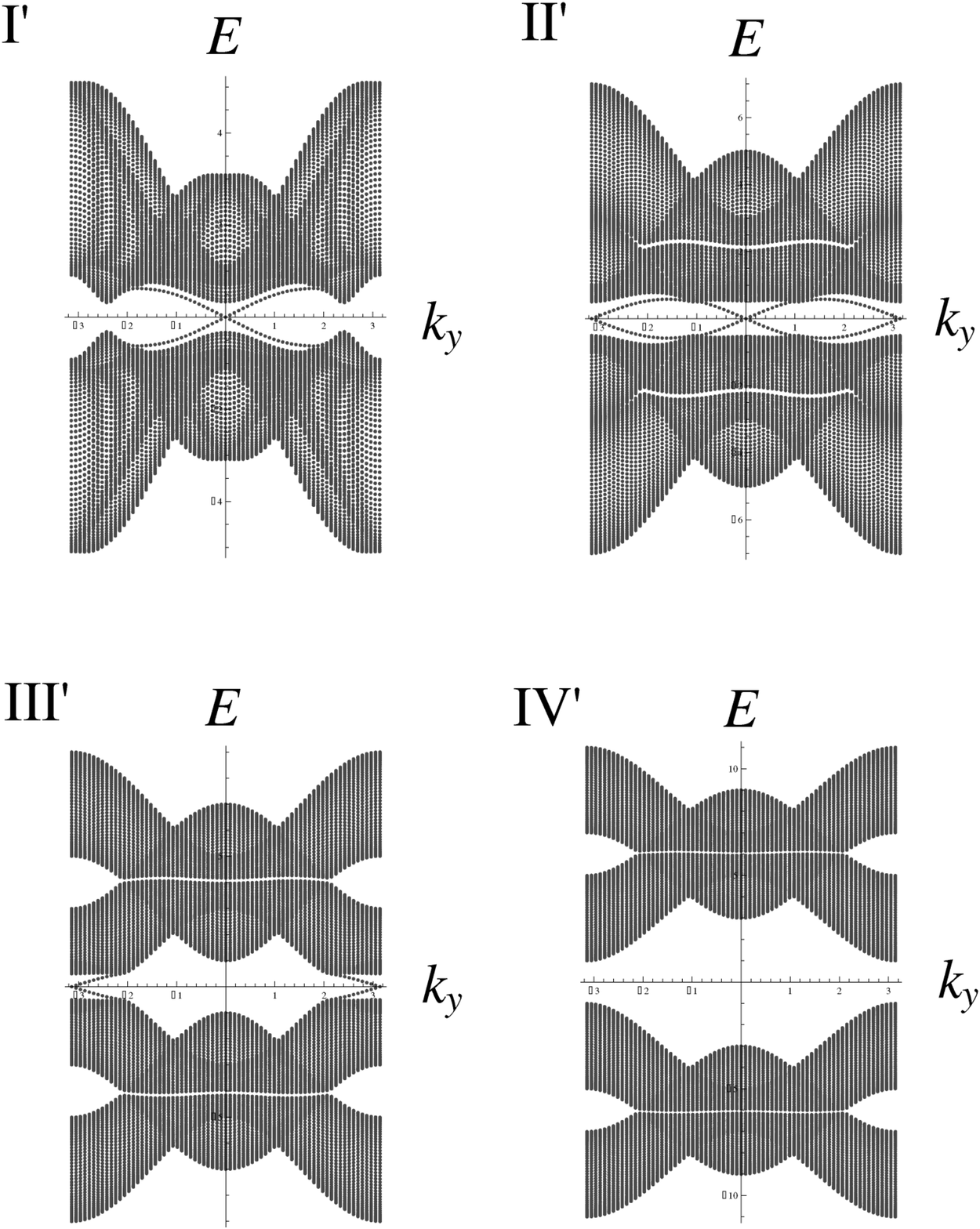}
\caption{The energy spectra of 2D NCS with edges at $i_x=0$ and
$i_x=50$ in phases I' II' III' and IV' in Fig.\ref{fig:phase} b).
Here $k_y$ denotes the momentum in the $y$-direction.
We take $t=1$, $\mu=-1$, $\alpha=0.6$, $\Delta_{\rm t}=0.6$ and
 $\Delta_{\rm s}=0.1$.
$H_z$ is I') $\mu_{\rm B}H_z=0$, II') $\mu_{\rm B}H_z=1$, III')
 $\mu_{\rm B}H_z=3$, IV') $\mu_{\rm B}H_z=5$.}
\label{fig:edgestatehzb} 
\end{figure}

In Figs.\ref{fig:edgestatehza} and \ref{fig:edgestatehzb}, we illustrate
the energy spectra of 2D NCS with edges for various topological phases.
All phases have a bulk gap, and some of them have gapless edge states
corresponding to the non-trivial topological numbers in Table
\ref{table:topologicalnumber}. 
It is found that for $k_y$ with nonzero $I(k_y)$ $(k_y=0,\pi)$, a zero
energy edge state always appears, and the number of zero energy edge states
coincides with the absolute value of $I(k_y)$. 
We also find that a phase with a non-zero $I_{\rm TKNN}$ has a edge state
with the total chirality $I_{\rm TKNN}$. 
These results are also consistent with the bulk-edge correspondence.

\begin{figure}
\includegraphics[width=10cm]{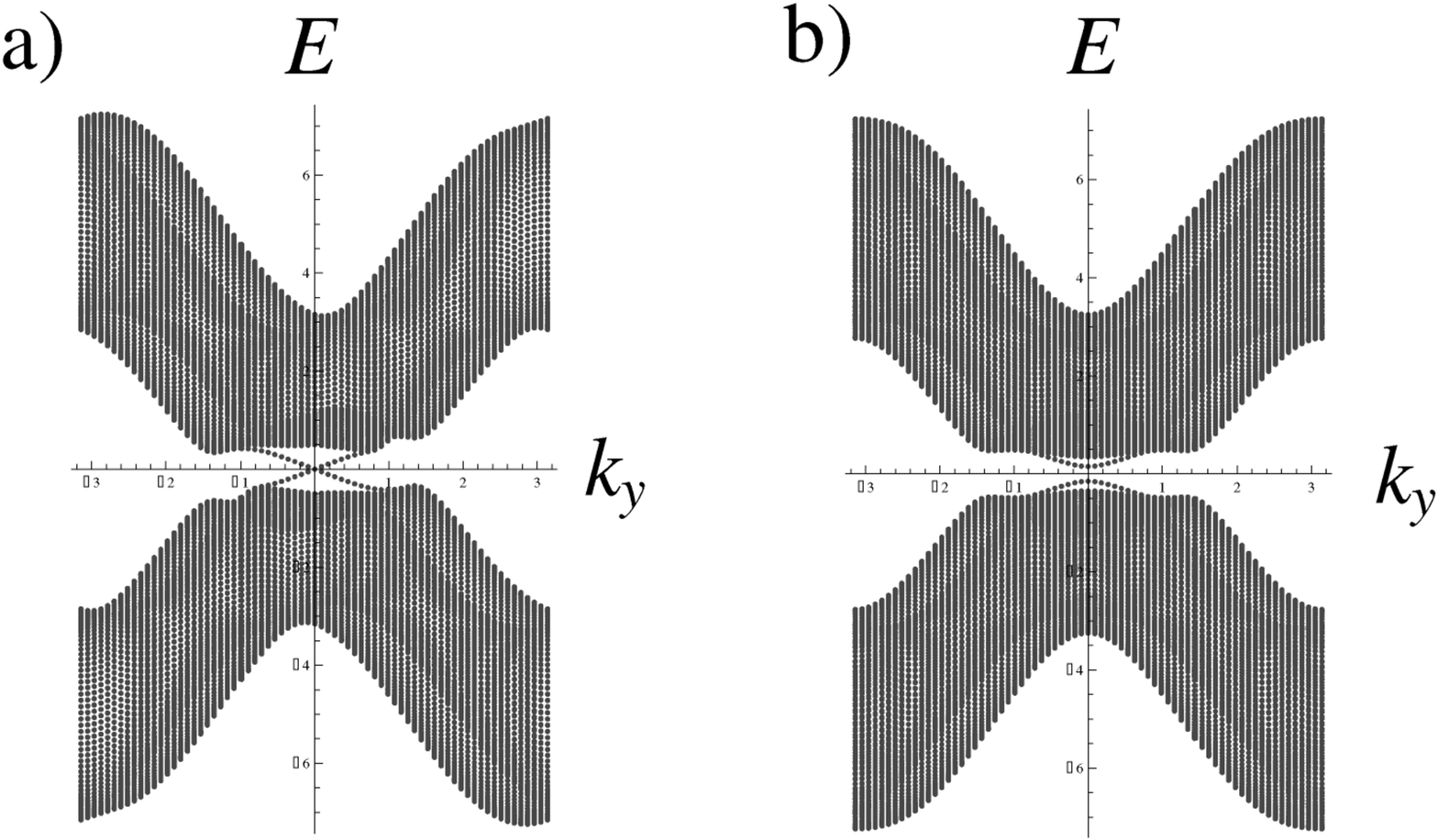}
\caption{The dependence of gapless edge states in the phase I on the
 direction of the magnetic field. 
We take $t=1$, $\mu=-3$, $\alpha=0.6$, $\Delta_{\rm t}=0.6$ and $\Delta_{\rm s}=0,1$
a) $\mu_{\rm B}H_x=0.15$, $\mu_{\rm B}H_y=0$ and $\mu_{\rm B}H_z=0$. b)
 $\mu_{\rm B}H_x=0$, $\mu_{\rm B}H_y=0.15$ and $\mu_{\rm B}H_z=0$.
 A similar dependence is obtained for gapless edge states in the phase I'.}
\label{fig:edgestatehy} 
\end{figure}

We also notice that 
the gapless edge states in the phases I and I'
are very sensitive to the direction of the applied magnetic field.
As seen in Fig.\ref{fig:edgestatehy},
while the gapless edge states are stable under a magnetic field in the
$x$-direction, they become unstable under a small magnetic field in the
$y$-direction.
This behavior is naturally understood by the sensitivity of the
definition of $I(k_y)$ to the direction of the magnetic field, which was
mentioned in the previous section:
For the phases I and I', the gapless edge states are ensured by 
$I(k_y)$, but in the case with non-zero $H_y$ its existence is no longer
protected since the winding number becomes ill-defined. 
As a result, the magnetic field $H_y$ along the edge causes a tiny gap
of the order
$O(\mu_{\rm B}H_y)$ for the edge states.

\subsection{Transport phenomena associated with  edge states}
\label{subsec:edgeexperiment}

\begin{figure}
\includegraphics[width=7cm]{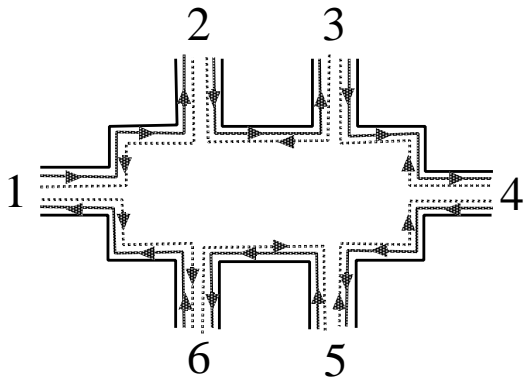}
\caption{Hall bar with six terminals used for the detection of edge states.
Solid and dotted gray lines with arrows represents propagating edge modes 
with different spin chirality corresponding to two SO split bands.
These modes propagate in opposite directions.}
\label{fig:hallbar} 
\end{figure}

The existence of gapless edge states revealed by the previous subsections
implies that transport phenomena associated with the edge states are possible
in analogy with the quantum Hall state and 
the ${\bm Z}_2$ topological insulator.
Here, we discuss such phenomena in NCS.
Transport properties inherent in edge states can be probed experimentally
by using the Hall bar geometry considered before for the detection of
edge states of 
the quantum (spin) Hall effect depicted 
in Fig. \ref{fig:hallbar} \cite{mce,wang,ber}.
Since our systems are superconductors, it is important for the experimental
detection to
discriminate between contributions from supercurrents
and currents carried by edge states.
A simple approach suitable for this purpose is 
to use thermal transport measurements.
To suppress contributions from the Bogoliubov quasiparticles in the bulk,
we assume that temperature is sufficiently lower than the
superconducting gap, $T\ll\Delta$,
and also the superconducting gap does not have nodes where
the gap vanishes.
The thermal conductance for heat currents  
is defined by
$G^{T}=I^{T}_{14}/(\Delta T)_{14}$ where
$I^{T}_{ij}$ is a thermal current between contacts $i$ and $j$ 
in Fig. \ref{fig:hallbar}, and
$(\Delta T)_{ij}$ is the temperature difference between these contacts.
In contrast to the conductance for electric currents,
the thermal conductance is not quantized but
depends on temperature $T$.
The $T$-dependence of $G^T$ governed by edge states obeys 
a power law $\propto T$,
which can be distinguished from contributions from
the bulk quasiparticles which decay like $\sim \exp(-\Delta/T)$.
Furthermore, as discussed in \cite{ber} in the case of the quantum spin
Hall effect, in a six terminal measurement, there is no temperature 
difference between contacts 2 and 3 (or 5 and 6), 
because the edge current is dissipationless.

A direct probe of spin Hall current carried by edge states may be also
possible by measuring magnetization due to spin accumulation at contacts
as discussed for the spin Hall effect \cite{zhang2}.
Bulk supercurrents carried by Cooper pairs do not contribute
to spin Hall currents even for a spin-triplet pairing state,
and hence
the spin Hall current between contacts 3 and 5 induced by
a longitudinal voltage or temperature difference between contacts 1 and 4
is governed by edge states.

A more remarkable effect due to edge states is the existence of
the non-local transport \cite{mce,wang}.
The non-local conductance is given by
a heat current between contacts 3 and 5 
divided by the temperature difference
between 2 and 6 in Fig.\ref{fig:hallbar}, 
$G^T_{\rm NL}=I^{T}_{35}/(\Delta T)_{26}$.
If contacts 3 and 5 are well separated from 2 and 6,
the nonzero $G^T_{NL}$ can not be explained by the bulk quasiparticles,
which provides a direct evidence for the existence of 
current-carrying edge states.

In the case with a sufficiently weak magnetic field,
the spin Hall effect still exists as long as the direction of
the magnetic field is perpendicular to the propagating direction of the
edge states, because of the accidental symmetry of the Rashba model 
as discussed in Secs. \ref{subsec:winding} and \ref{subsec:withh}. 
If the magnetic field is tilted, 
and the field component along
the propagating direction is nonzero, a gap opens in the energy spectrum of
the edge states;
this leads to
the suppression of the spin Hall current which can be observed 
as a drop in the temperature difference between contacts.

The gapless edge states are also observed
experimentally as a zero bias peak of tunneling current \cite{IHSYMTS07}.
Our results suggest that structures of the zero bias peak is very sensitive to the
direction of the magnetic field.
Under a magnetic field perpendicular to the edge direction, the zero
bias peak is observed, but under a magnetic field along the
edge, a splitting of the conduction peak is observed due to a tiny gap
of the gapless edge states.

The transport properties considered above are experimentally observable
for 2D Rashba NCS with $\Delta_{\rm t}>\Delta_{\rm s}$ 
which do not possess gap-nodes.

 \section{Majorana zero energy modes in vortex cores 
and the non-Abelian statistics}
\label{sec:vortex}

In this section, we explore zero energy states in vortex cores for 2D NCS
on the basis of both numerical and analytical approaches.
We study the case with purely $p$-wave paring and the case with the
admixture of $s$-wave and $p$-wave parings, respectively.
It is assumed that no node exists in the superconducting gap 
because the existence of a full
energy gap in the bulk is crucial for the stability of topological phases.
As mentioned in the previous sections, 
as long as the gap of $p$-wave pairing is larger than that
of $s$-wave pairing, the state is topologically equivalent to
the purely $p$-wave pairing state, exhibiting topological
nontriviality.
In the following, 
we confirm this topological argument
by obtaining explicitly the zero energy solutions of vortex cores.
The condition for the non-Abelian statistics of vortices is 
also clarified on the basis of the explicit solutions.

\subsection{Numerical analysis of BdG equations}

In this subsection, we analyze the low energy states in a vortex core 
of NCS by using numerical methods.
For this purpose, we consider 
the following two dimensional tight-binding model for
a Rashba superconductor with the admixture of $s$-wave and $p$-wave pairings,
\begin{eqnarray}
\mathcal{H}&=&-t\sum_{\langle {\bm i},{\bm j}\rangle ,\sigma}
c^{\dagger}_{{\bm i}\sigma}c_{{\bm j}\sigma}
-\mu\sum_{{\bm i},\sigma}c^{\dagger}_{{\bm i}\sigma}c_{{\bm i}\sigma}
-\mu_{\rm B}H_z \sum_{{\bm i},\sigma,\sigma'}(\sigma_z)_{\sigma\sigma'}
c_{{\bm i}\sigma}^{\dagger}c_{{\bm i}\sigma'}
-i\frac{\alpha}{2}\sum_{\langle {\bm i},{\bm j}\rangle ,\sigma,\sigma'}
({\bm \sigma}_{\sigma\sigma'}
\times\hat{{\bm r}}_{{\bm i}{\bm j}})_z
c^{\dagger}_{{\bm i}\sigma}c_{{\bm j}\sigma'}
 \nonumber \\
&&+\Delta_{\rm s}\sum_{\bm i}(e^{i\phi_{\bm i}}
 c^{\dagger}_{{\bm i}\uparrow}c^{\dagger}_{{\bm i}\downarrow}+{\rm h.c.}) 
\nonumber \\
&&-\frac{1}{2}\Delta_{\rm t}\sum_{\bm i}(e^{i\phi_{{\bm i}+\frac{\hat{x}}{2}}} 
c^{\dagger}_{{\bm i}\uparrow}c^{\dagger}_{{\bm i}+\hat{x}\uparrow}
-ie^{i\phi_{{\bm i}+\frac{\hat{y}}{2}}} c^{\dagger}_{{\bm i}\uparrow}
c^{\dagger}_{{\bm i}+\hat{y}\uparrow} 
+e^{i\phi_{{\bm i}+\frac{\hat{x}}{2}}} 
c^{\dagger}_{{\bm i}\downarrow}c^{\dagger}_{{\bm i}+\hat{x}\downarrow}
+ie^{i\phi_{{\bm i}+\frac{\hat{y}}{2}}} 
c^{\dagger}_{{\bm i}\downarrow}c^{\dagger}_{{\bm
i}+\hat{y}\downarrow}+{\rm h.c.}).
\label{tbham}
\end{eqnarray}
Here the second and third lines of the right-hand side are, respectively,
the $s$-wave and $p$-wave pairing terms, and 
a vortex located on the center of the system is incorporated into
the phase $\phi_{\bm i}$ of the gap functions.
To obtain the vortex core states for $\mathcal{H}$,
we introduce the Bogoliubov quasiparticle operator,
\begin{eqnarray}
\gamma^{\dagger}=\sum_{\bm i}[u_{\uparrow}({\bm i})
c^{\dagger}_{{\bm i}\uparrow}+
u_{\downarrow}({\bm i})c^{\dagger}_{{\bm i}\downarrow}+
v_{\uparrow}({\bm i})c_{{\bm i}\uparrow}+
v_{\downarrow}({\bm i})c_{{\bm i}\downarrow}].
\end{eqnarray}
The Bogoliubov-de Gennes (BdG) equations is derived from
the relation $[\mathcal{H},\gamma^{\dagger}]=E\gamma^{\dagger}$.
We solve the BdG equations numerically, and calculate
the density profile of quasiparticles for low energy states.

\begin{figure}
\includegraphics[width=15cm]{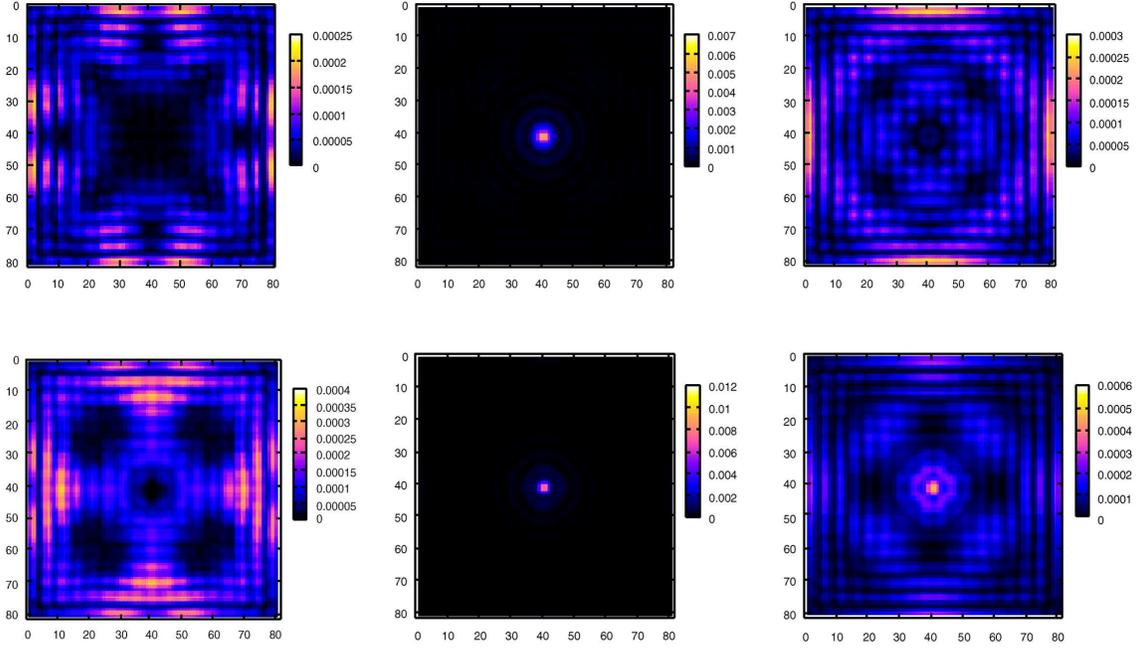}
\caption{\label{Fig3-1} The density of quasiparticles for the purely $p$ wave
state plotted on the $81\times 81$ $xy$-plane
for some low energies $E$. 
The top and bottom panels are, respectively, 
the plots of $|u_{\uparrow}({\bm r})|^2
+|u_{\downarrow}({\bm r})|^2$ and 
$|v_{\uparrow}({\bm r})|^2
+|v_{\downarrow}({\bm r})|^2$ for
$\alpha=t$, $\Delta_{\rm t}=0.05t$, $\mu_{\rm B}H_z=0$, $\mu=-3.75t$.
$E=4.87\times 10^{-4}t$ (left), 
$E=7.67\times 10^{-4}t$ (middle),
$E=1.023\times 10^{-3}t$ (right).
For $E=4.87\times 10^{-4}t$ and $E=1.023\times 10^{-3}t$,
edge states appear.}
\end{figure}
\begin{figure}
\includegraphics[width=15cm]{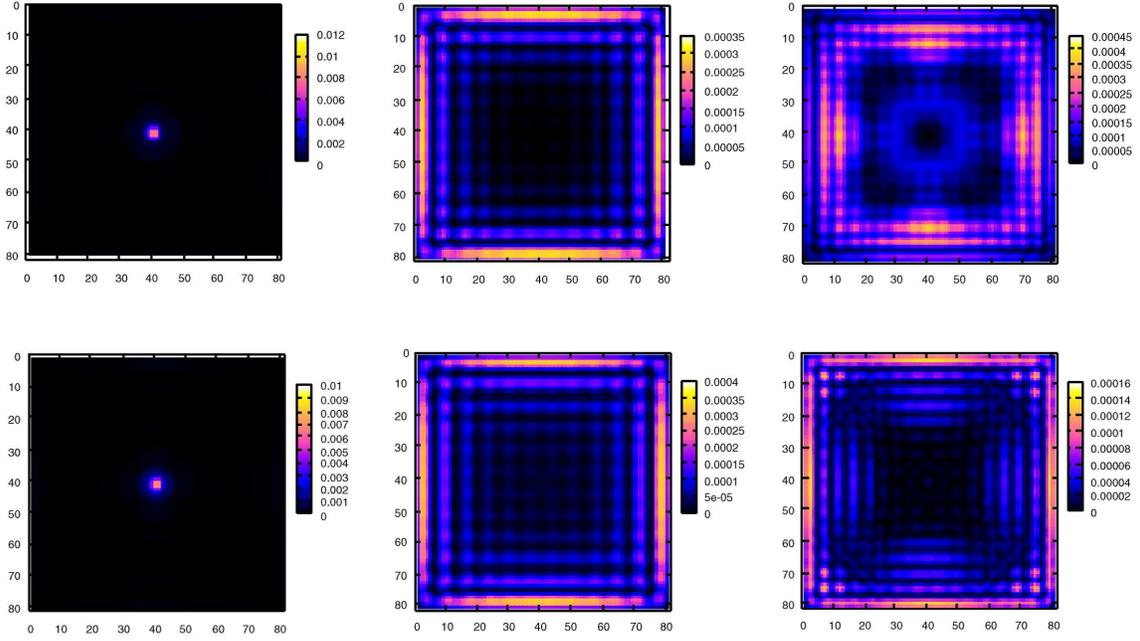}
\caption{\label{Fig3-2} The density of quasiparticles
for the purely $p$ wave state plotted
on the $81\times 81$ $xy$-plane
for low energies.
The top and bottom panels are, respectively, 
the plots of $|u_{\uparrow}({\bm r})|^2
+|u_{\downarrow}({\bm r})|^2$ and 
$|v_{\uparrow}({\bm r})|^2
+|v_{\downarrow}({\bm r})|^2$.
$\alpha=t$, $\Delta_{\rm t}=0.05t$, $\mu_{\rm B}H_z=0.04t$, $\mu=-4.0t$.
$E=0.00123t$ (left), 
$E=0.0182t$ (middle),
$E=0.0315t$ (right).
Edge states appear for $E=0.0182t$ and $E=0.0315t$. 
}
\end{figure}

We, first, consider the case with a purely $p$-wave pairing.
In Fig.\ref{Fig3-1}, the density of quasiparticles plotted
on the $xy$-plane is shown.
There are low energy vortex core states.
The lowest energy for the vortex core state is
$E=0.000767t$.
For our choice of parameters in this calculation,
$\Delta_{\rm t}^2/E_F\sim 0.005$.
Thus, the vortex core state has much smaller energy than
that of the conventional Caroli-de Gennes-Matricon mode.
Furthermore, we find edge states with low energies
as shown in Fig.\ref{Fig3-1}. 
The existence of low energy edge states
is a concomitant of zero energy modes in vortex cores,
and characterizes topological nature of the state.
Therefore, we conclude that the zero energy vortex core states
exist.

The zero energy vortex core states and the low energy edge states
survive even when a Zeeman field along the $z$-direction is applied.
The calculated results in this case is shown in Fig.\ref{Fig3-2}.

\begin{figure}
\includegraphics[width=15cm]{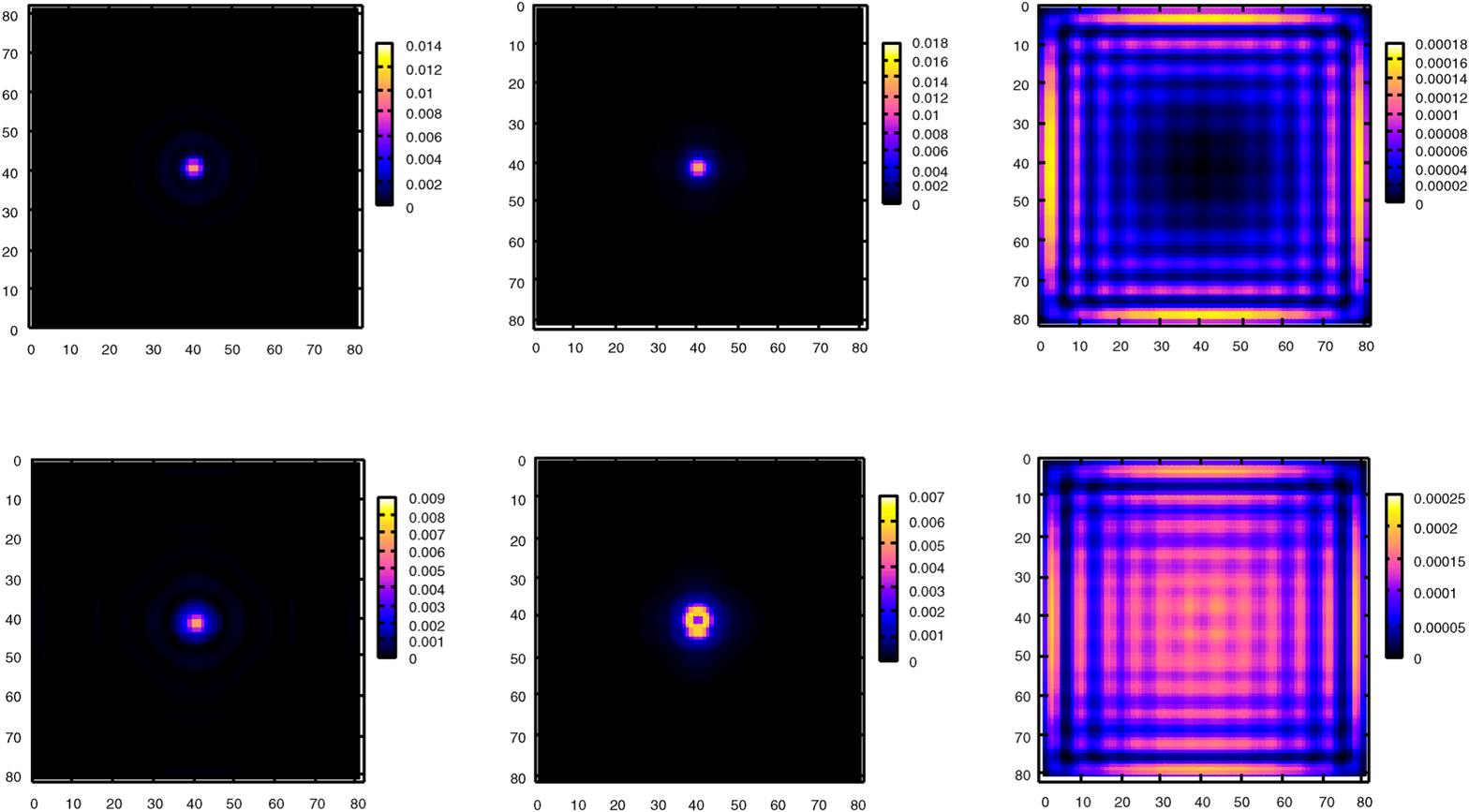}
\caption{\label{Fig3-3} The density of quasiparticles
for the $s+p$ wave state plotted on the $81\times 81$ $xy$-plane
for low energies.
The top and bottom panels are, respectively, 
the plots of  $|u_{\uparrow}({\bm r})|^2
+|u_{\downarrow}({\bm r})|^2$ and 
$|v_{\uparrow}({\bm r})|^2
+|v_{\downarrow}({\bm r})|^2$. 
$\alpha=t$, $\Delta_{\rm t}=0.05t$, $\Delta_{\rm s}=-0.02t$, $\mu_{\rm B}H_z=0$, $\mu=-4.0t$.
$E=0.0023t$ (left), 
$E=0.039t$ (middle),
$E=0.0434t$ (right).
Edge states appear for $E=0.0434t$.}
\end{figure}

We, next, consider the case with the admixture of $s$-wave pairing
and $p$-wave pairing.
The numerical results shown in Fig.\ref{Fig3-3} demonstrate that
the topological phase with gapless edge modes and zero energy vortex
core states is stable as long as 
the gap of $p$-wave pairing is larger than that of
$s$-wave pairing.
As the magnitude of $|\Delta_{\rm s}|$ approaches $|\Delta_{\rm t}|$,
the gap in one of two SO split bands $|\Delta_{\rm t}|-|\Delta_{\rm s}|$
decreases,
and the spectrum gap in the vortex core becomes smaller, which
makes it difficult to clarify the existence of zero energy vortex core
states numerically.
The numerical approach in this section is applicable only when
$|\Delta_{\rm t}|-|\Delta_{\rm s}|$ is not much smaller than $|\Delta_{\rm t}|$.

To discuss whether the non-Abelian statistics of vortices is possible or not,
we need to examine the degeneracy of Majorana modes.
The numerical analysis presented in this subsection is not suitable
for this purpose because of the limitation of the system size
used in the numerics.
To attack this issue, we exploit an analytical method in
Sec. \ref{subsec:nonabelian}.

\subsection{A comment on the use of the chirality basis
for the issues of vortex core states}
\label{subsec:chiralitybasis}

Here, we discuss the limitation of the use of
the chirality basis for the issues
of vortex core states.
For concreteness, we consider a noncentrosymmetric $s$-wave superconductor.
In the absence of Zeeman fields,
in homogeneous bulk systems without vortices,
the unitary transformation $(a_{{\bm k}+},a_{{\bm k}-})^{T}=
\hat{U}({\bm k})(c_{{\bm k}\uparrow},c_{{\bm k}\downarrow})^{T}$ with
\begin{eqnarray}
\hat{U}({\bm k})=
\frac{1}{\sqrt{2}}\left(
\begin{array}{cc}
1  &   e^{-i\theta({\bm k})} \\
-e^{i\theta({\bm k})} & 1 
\end{array}
\right), \label{unita}
\end{eqnarray}
which diagonalizes the asymmetric SO interaction
in the spin space transforms the $s$-wave pairing terms into
the intra-band pairing terms in the chirality basis,
\begin{eqnarray}
\sum_{\nu=\pm}\sum_{\bm k}\Delta_{\rm s}e^{\nu i\theta({\bm k})}
a_{{\bm k}\nu}^{\dagger}a_{-{\bm k}\nu}^{\dagger}+{\rm h.c.},
\end{eqnarray}
where $e^{\pm i\theta({\bm k})}\equiv -(\mathcal{L}_{0x}\pm i \mathcal{L}_{0y})
/\sqrt{\mathcal{L}_{0x}^2+\mathcal{L}_{0y}^2}$ is an odd-parity phase factor.
The above expression is similar to the pairing term of $p+ip$ superconductors.
Thus, one may expect that zero energy vortex core modes exist when vortices
are introduced into the system.
However, this is not true.
The unitary transformation (\ref{unita}) is derived for 
spatially homogeneous systems, and generally 
its use is not allowed in the case with vortices.
In the case that quasiclassical approximation is justified, and
the center of mass coordinate of Cooper pairs 
which characterizes inhomogeneity of the system 
is approximately commutative with the momentum operator ${\bm k}$ and the unitary
transformation $\hat{U}({\bm k})$,
the chirality basis can be used for the description of the vortex state,
as done in \cite{hayashi,kato,fuji2}.
However, to examine whether the zero energy mode exists or not in 
a vortex core, one needs to go beyond the quasiclassical approximation; i.e.
the unitary transformation for homogeneous systems (\ref{unita})
is not applicable to the issue that we are concerned with here.
Missing this consideration in \cite{fuji1} leads to an incorrect 
conclusion that zero energy vortex core states exist even for purely $s$-wave
NCS.
For the clarification of the existence of zero energy modes in vortex cores, 
a powerful and reliable approach is to exploit
the method developed by Tewari, Sarma and Lee which is analogous to
the Jackiw-Rebbi's index theorem \cite{tewari,jack}.
We consider this issue in the subsequent sections.

\subsection{Majorana zero energy modes in vortex cores: Index theorem}
\label{subsec:index}

In this subsection, we develop an analytical approach for 
zero energy vortex core states which is a generalization of
the index theorem for $p+ip$ superconductors obtained in \cite{tewari} to 
the case of noncentrosymmetric systems.
We consider the model for a Rashba superconductor in two dimensions
with a vortex located on the center of the system.
The Hamiltonian is
\begin{eqnarray}
&&\mathcal{H}=\mathcal{H}_{\rm K}+\mathcal{H}_{\rm SO}+\mathcal{H}_{\rm
 Ps}+\mathcal{H}_{\rm Pt},
\nonumber\\
&&\mathcal{H}_{\rm K}=\sum_{{\bm k},\sigma}\xi_{{\bm k}\sigma}
c^{\dagger}_{{\bm k}\sigma}c_{{\bm k}\sigma},
\nonumber\\
&&\mathcal{H}_{\rm SO}=\alpha\sum\mbox{\boldmath $\mathcal{L}$}({\bm k})
\cdot{\bm \sigma}_{\sigma\sigma'}
c^{\dagger}_{{\bm k}\sigma}c_{{\bm k}\sigma'}, \quad
\mbox{\boldmath $\mathcal{L}$}({\bm k})=(k_y,-k_x,0),
\nonumber\\
&&\mathcal{H}_{\rm Ps}=\Delta_{\rm s}\int d^2{\bm R}
d^2{\bm r}e^{i\phi_R}h_s({\bm R})g_s({\bm r})
c^{\dagger}_{{\bm R}+{\bm r} \uparrow}
c^{\dagger}_{{\bm R}-{\bm r} \downarrow}+{\rm h.c.},
\nonumber\\
&&\mathcal{H}_{\rm Pt}=-\frac{1}{2}\Delta_{\rm t}\int d^2{\bm R}
d^2{\bm r}e^{i\phi_R}h_s({\bm R})
[g_{\uparrow}({\bm r})
c^{\dagger}_{{\bm R}+{\bm r} \uparrow}
c^{\dagger}_{{\bm R}-{\bm r} \uparrow}+
g_{\downarrow}({\bm r})
c^{\dagger}_{{\bm R}+{\bm r} \downarrow}
c^{\dagger}_{{\bm R}-{\bm r} \downarrow}]+{\rm h.c.},
\end{eqnarray}
where the band dispersion in the case with the Zeeman field $H_z$ is
$\xi_{{\bm k}\sigma}=\epsilon_{\bm k}-\mu-\sigma \mu_{\rm B}H_z$ with $\epsilon_{\bm
k}={\bm k}^2/2m$ and 
$\mu$ a chemical potential.
$\mathcal{H}_{\rm SO}$ is the Rashba SO interaction with the coupling constant
$\alpha$.
$\mathcal{H}_{\rm Ps}$ and $\mathcal{H}_{\rm Pt}$ are, respectively,
the pairing interaction for spin-singlet and spin-triplet channels.
The center of mass coordinate and the relative coordinate for Cooper pairs
are, respectively, represented by
${\bm R}$
and ${\bm r}$. 
The phase $\phi_R$ is due to the vortex at ${\bm R}=0$.
The suppression of the superconducting order parameter 
in the vicinity of the vortex is incorporated into the functions
$h_a({\bm R})=1-e^{-|{\bm R}|/\xi_a}$ with $a={\rm s}$ for the
spin-singlet pair and $a={\rm t}$ for
the spin-triplet pair.
$g_{\rm s}({\bm r})$, $g_{\uparrow}({\bm r})$, and $g_{\downarrow}({\bm r})$ are
the structure functions corresponding to the pairing symmetry.
For simplicity, in the following, we consider the case with $s$-wave pairing for
the spin-singlet channel and the $p$-wave pairing for the spin-triplet channel.
Then, the Fourier transforms for the gap structure functions are
$g_{\rm s}({\bm k})=1$, $g_{\uparrow}({\bm k})=-i(k_x-ik_y)$ 
and $g_{\downarrow}({\bm k})=-i(k_x+ik_y)$.
Here the form of the ${\bm d}$-vector for triplet pairs are determined so as
to be consistent with the Rashba SO interaction.
To discuss the zero energy state in the vortex core, we follow the approach 
developed in \cite{tewari}, and use the angular momentum representation of
the electron operators,
\begin{eqnarray}
c_{{\bm k}\sigma}
=\frac{1}{\sqrt{2\pi k}}\sum_{m=-\infty}^{\infty}c_{m,k\sigma}e^{i m\phi_k},
\end{eqnarray}
where $k=|{\bm k}|$ and $\phi_k$ is the azimuthal angle of ${\bm k}$.
$c_{m,k\sigma}$ satisfies the anticommutation relation 
$\{c_{m,k\sigma},c^{\dagger}_{n,p\sigma'}\}=\delta_{mn}\delta_{\sigma\sigma'}\delta(k-p)$.
In this representation, the Hamiltonian is rewritten into
the following form.
\begin{eqnarray}
\mathcal{H}_{\rm K}=\sum_m\int \frac{dk}{(2\pi)^2}\xi_{k\sigma}
c^{\dagger}_{m,k\sigma}c_{m,k\sigma},
\label{hama1}
\end{eqnarray}
\begin{eqnarray}
\mathcal{H}_{\rm SO}=\alpha\sum_m\int \frac{dk}{(2\pi)^2} 
ik[c^{\dagger}_{m,k\uparrow}c_{m+1,k\downarrow}
-c^{\dagger}_{m,k\downarrow}c_{m-1,k\uparrow}]
\label{hama2}
\end{eqnarray}
\begin{eqnarray}
\mathcal{H}_{\rm Ps}=-i\Delta_{\rm s}\int dk dp
\sum_m u_m(k,p)\sqrt{kp}
[kc^{\dagger}_{1-m,k\uparrow}c^{\dagger}_{m,p\downarrow}
+pc^{\dagger}_{-m,k\uparrow}c^{\dagger}_{m+1,p\downarrow}]
+{\rm h.c.}, 
\label{hama3}
\end{eqnarray}
\begin{eqnarray}
\mathcal{H}_{\rm Pt}=\Delta_{\rm t}\int dk dp\sum_m[\{u_m(k,p)k-u_{m+1}(k,p)p\}k\sqrt{kp}
c^{\dagger}_{-m,k\uparrow}c^{\dagger}_{m,p\uparrow}
+u_m(k,p)k^2\sqrt{kp}
c^{\dagger}_{2-m,k\downarrow}c^{\dagger}_{m,p\downarrow}]+{\rm h.c.}.
\label{hama4}
\end{eqnarray}
Here $u_m(k,p)$ is the Fourier transform of $1/|{\bm k}+{\bm p}|^3$.
\begin{eqnarray}
\frac{1}{|{\bm k}+{\bm p}|^3}
=\sum_mu_m(k,p)e^{-im(\phi_{\bm k}-\phi_{\bm p})}.
\end{eqnarray}
In Eqs.(\ref{hama1}),(\ref{hama2}), (\ref{hama3}) and (\ref{hama4}),
the $m=0$ mode with up spin and the $m=1$ mode with down spin are decoupled from
other modes.
The pairing interaction terms for these modes are
\begin{eqnarray}
&&\mathcal{H}_{\rm Ps}^{(01)}=-i\Delta_{\rm s}\int dk dp\sqrt{kp}[u_1(k,p)k+
u_0(k,p)p]
c^{\dagger}_{0,k\uparrow}c^{\dagger}_{1,p\downarrow}
+{\rm h.c.},
\nonumber\\
&&\mathcal{H}_{\rm Pt}^{(01)}=\Delta_{\rm t}\int dk dp[\{u_0(k,p)k-u_{1}(k,p)p\}k\sqrt{kp}
c^{\dagger}_{0,k\uparrow}c^{\dagger}_{0,p\uparrow}
+u_1(k,p)k^2\sqrt{kp}
c^{\dagger}_{1,k\downarrow}c^{\dagger}_{1,p\downarrow}]+{\rm h.c.}.
\end{eqnarray}
In $\mathcal{H}_{\rm Pt}^{(01)}$, 
the term with $u_1(k,p)$ for the $\uparrow\uparrow$ pairs 
vanishes because
of the symmetric property $u_1(k,p)=u_1(p,k)$.
Thus, we have
\begin{eqnarray}
\mathcal{H}^{(01)}_{\rm Pt}=\Delta_t\int dk dp\sqrt{kp}k^2[u_0(k,p)
c^{\dagger}_{0,k\uparrow}c^{\dagger}_{0,p\uparrow}+
u_1(k,p)c^{\dagger}_{1,k\downarrow}c^{\dagger}_{1,p\downarrow}]+{\rm h.c.}.
\end{eqnarray}

\subsubsection{Purely $p$-wave pairing}

We, first, consider the purely triplet case, i.e. $\Delta_{\rm s}=0$.
Then, the Hamiltonian for the $m=0$ mode with up spin and the $m=1$ mode
with down spin is
\begin{eqnarray}
\mathcal{H}^{(01)}&=&\int \frac{dk}{(2\pi)^2}
[\xi_{k\uparrow}c^{\dagger}_{0,k\uparrow}c_{0,k\uparrow}
+\xi_{k\downarrow}c^{\dagger}_{1,k\downarrow}c_{1,k\downarrow}] \nonumber \\
&+&i\alpha\int \frac{dk}{(2\pi)^2} k[c^{\dagger}_{0,k\uparrow}c_{1,k\downarrow}
-c^{\dagger}_{1,k\downarrow}c_{0,k\uparrow}]  \nonumber \\
&-&\int dk dp 
[\tilde{\Delta}_{{\rm t}0}(k,p)c^{\dagger}_{0,k\uparrow}c^{\dagger}_{0,p\uparrow}
- \tilde{\Delta}_{{\rm t}1}
c^{\dagger}_{1,k\downarrow}c^{\dagger}_{1,p\downarrow}+{\rm h.c.}],
\label{ham01}
\end{eqnarray}
with $\tilde{\Delta}_{\rm t0}(k,p)=\Delta_{\rm t}\sqrt{kp}(k^2-p^2)u_0(k,p)/2$, and
$\tilde{\Delta}_{\rm t1}(k,p)=-\Delta_{\rm t}\sqrt{kp}(k^2-p^2)u_1(k,p)/2$.
Here we have antisymmetrized the spin-triplet pairing terms.
In the Fourier coefficients,
\begin{eqnarray}
u_0(k,p)=\int \frac{d(\phi_{\bm k}-\phi_{\bm p})}{2\pi}\frac{1}{|{\bm k}+
{\bm p}|^3},
\label{u0f}
\end{eqnarray}
\begin{eqnarray}
u_1(k,p)=\int \frac{d(\phi_{\bm k}-\phi_{\bm p})}
{2\pi}\frac{e^{i(\phi_{\bm k}-\phi_{\bm p})}}
{|{\bm k}+{\bm p}|^3},
\label{u1f}
\end{eqnarray}
the most dominant contributions are from $\phi_{\bm k}-\phi_{\bm p}\sim\pm\pi$, which implies
$u_1(k,p)<0$ and $u_0(k,p)>0$.
By using the unitary transformation
\begin{eqnarray}
c_{0,k\uparrow}=\zeta_{+}(k)a_{k+}+i\zeta_{-}(k)a_{k-},
\label{ut1}
\end{eqnarray}
\begin{eqnarray}
c_{1,k\downarrow}=i\zeta_{-}(k)a_{k+}+\zeta_{+}(k)a_{k-},
\label{ut2}
\end{eqnarray}
with 
\begin{eqnarray}
\zeta_{\pm}(k)=\sqrt{\frac{1}{2}
\left(1\pm \frac{H_z}{\sqrt{\alpha^2k^2+H_z^2}}\right)},
\end{eqnarray}
we rewrite the Hamiltonian (\ref{ham01}) as,
\begin{eqnarray}
\mathcal{H}^{(01)}&=&\sum_{\nu=\pm}\bigl[
\int \frac{dk}{(2\pi)^2}\xi_{k}^{(\nu)}a^{\dagger}_{k\nu}a_{k\nu}
+\int dk \int dp \{\tilde{\Delta}_{\nu}(k,p)a^{\dagger}_{k\nu}a^{\dagger}_{p\nu}
+{\rm h.c.}\}\bigr] \nonumber \\
&+&\int dk\int dp [\tilde{\Delta}_{2+}(k,p)a^{\dagger}_{k+}a^{\dagger}_{p-}+
\tilde{\Delta}_{2-}(k,p)a^{\dagger}_{k-}a^{\dagger}_{p+}+{\rm h.c.}],
\label{ham01t}
\end{eqnarray}
where 
\begin{eqnarray}
\tilde{\Delta}_{\pm}(k,p)=\pm\frac{\tilde{\Delta}_{\rm t0}
+\tilde{\Delta}_{{\rm t}1}}{2}r_{+}(k,p)+
\frac{\tilde{\Delta}_{\rm t0}
-\tilde{\Delta}_{{\rm t}1}}{2}r_{-}(k,p),
\label{delpm}
\end{eqnarray}
\begin{eqnarray}
\tilde{\Delta}_{2\pm}(k,p)=\mp i\frac{\tilde{\Delta}_{\rm t0}
+\tilde{\Delta}_{{\rm t}1}}{2}s_{-}(k,p)-
i\frac{\tilde{\Delta}_{\rm t0}
-\tilde{\Delta}_{{\rm t}1}}{2}s_{+}(k,p),
\label{del2}
\end{eqnarray}
with $r_{\pm}(k,p)=\zeta_{+}(k)\zeta_{+}(p)\pm \zeta_{-}(k)\zeta_{-}(p)$,
$s_{\pm}(k,p)=\zeta_{+}(k)\zeta_{-}(p)\pm \zeta_{-}(k)\zeta_{+}(p)$,
and $\xi_{k}^{(\nu)}=\epsilon_k-\mu-\nu\sqrt{\alpha^2k^2+(\mu_{\rm B}H_z)^2}$.
We concentrate on low energy excitations in the vicinity of the Fermi surface, and
write $k$ and $p$ as $k=k_{F\pm}+q$ and $p=k_{F\pm}+q'$ where
$k_{F+}$ and $k_{F-}$ are, respectively, 
the Fermi momenta for the two SO split bands, and $|q|,|q'|\ll k_{F\pm}$.
In Eq.(\ref{del2}), the first term is of the order $O((q-q')^2)$, because
$\tilde{\Delta}_{\rm t0,1}(k,p)$ and $s_{-}(k,p)$ are antisymmetric with respect to
the exchange of $k$ and $p$.
We can neglect this term compared to other terms up to $O(q-q')$. 
Furthermore, from Eqs.(\ref{u0f}) and (\ref{u1f}),
we see that 
$|\tilde{\Delta}_{\rm t0}-\tilde{\Delta}_{\rm t1}|
/|\tilde{\Delta}_{\rm t0}+\tilde{\Delta}_{\rm t1}|\sim O(1/R_{\rm c}^2)$
where $R_{\rm c}$ is 
the system size,
and in the limit of $R_{\rm c}\rightarrow \infty$,
the second term of (\ref{del2}) is negligible compared
to the first term of (\ref{delpm}).
As a result, the inter-band pairing terms can be neglected, and 
the Hamiltonian (\ref{ham01t}) is decoupled into
two parts corresponding, respectively, to contributions
from the two SO split bands, $\nu=+$ and $-$.
$\tilde{\Delta}_{\pm}(k,p)\equiv A_{\pm}(q-q')$ 
must be odd in $q-q'$, since, otherwise, the pairing term 
in Eq.(\ref{ham01t}) vanishes.
Then, the Fourier transforms of $A_{\pm}(q-q')$ denoted as $\mp i m_{\pm}(x)$ 
are odd functions
of $x$. Here $m_{\pm}(x)$ is real.

Linearizing the band dispersion $\xi_{k\nu}$ around the Fermi momentum, 
and expressing the operator for the Bogoliubov quasiparticle as,
\begin{eqnarray}
\gamma^{\dagger}_{\nu}=\int dx
[\eta_{1\nu}a^{\dagger}_{\nu}(x)+\eta_{2\nu}a_{\nu}(-x)],
\label{quasi}
\end{eqnarray}
we obtain the Bogoliubov-de-Gennes (BdG) equations from the relation
$[\mathcal{H}^{(01)},\gamma^{\dagger}]=E\gamma^{\dagger}$, 
\begin{eqnarray}
-iv_{\nu}\sigma_z\partial_x\eta_{\nu}(x)+\nu m_{\nu}(x)\sigma_y\eta_{\nu}(x)=E
\eta_{\nu}(x),
\label{bdg}
\end{eqnarray}
with $\nu=\pm$ corresponding to the two SO split bands and the Fermi
velocities $v_{\nu}$.
Here $\eta_{\pm}^T(x)=(\eta_{1\pm},\eta_{2\pm})$.
The BdG equations have a zero energy solution with $E=0$ 
for each bands:
When $m_{\pm}(x)$ satisfies $m_{\pm}(x)={\rm sgn}(x)|m_{\pm}(x)|$, the
zero energy solutions are
\begin{eqnarray}
\eta_{+}^T(x)=e^{-\frac{1}{v_{+}}\int^x_0  m_{+}(y)dy}\frac{1}{\sqrt{2}}(1,1),
\label{sol1}
\end{eqnarray}
\begin{eqnarray}
\eta_{-}^T(x)=e^{-\frac{1}{v_{+}}\int^x_0  m_{-}(y)dy}\frac{i}{\sqrt{2}}(1,-1).
\label{sol2}
\end{eqnarray}
For these solutions (\ref{sol1}) and (\ref{sol2}), the quasiparticle 
operator (\ref{quasi}) satisfies $\gamma_{\nu}^{\dagger}=\gamma_{\nu}$, 
and thus there are two Majorana fermion modes
corresponding to the two bands.

It is noted that these two Majorana modes are stable against
the magnetic field along $H_z$ in accordance with the existence of two gapless
edge modes in this case as discussed in Sec. \ref{subsec:withh}.
This phase is topologically equivalent to
the phase I or I' in Table \ref{table:topologicalnumber}.

\subsubsection{$s+p$ wave pairing}

Now we consider the case with the admixture of the spin singlet pairs
and spin triplet pairs.
The Hamiltonian for the $m=0$ mode with up spin and the $m=1$ mode with down spin
is
\begin{eqnarray}
\mathcal{H}^{(01)}&=&\int \frac{dk}{(2\pi)^2}
[\xi_{k\uparrow}c^{\dagger}_{0,k\uparrow}c_{0,k\uparrow}
+\xi_{k\downarrow}c^{\dagger}_{1,k\downarrow}c_{1,k\downarrow}] \nonumber \\
&+&i\alpha\int \frac{dk}{(2\pi)^2} k[c^{\dagger}_{0,k\uparrow}c_{1,k\downarrow}
-c^{\dagger}_{1,k\downarrow}c_{0,k\uparrow}]  \nonumber \\
&-&i\int dk dp \tilde{\Delta}_{\rm s}(k,p)[ 
c^{\dagger}_{0,k\uparrow}c^{\dagger}_{1,p\downarrow}
+{\rm h.c.}] \nonumber \\ 
&+&\int dk dp 
[\tilde{\Delta}_{\rm t0}(k,p)c^{\dagger}_{0,k\uparrow}c^{\dagger}_{0,p\uparrow}
- \tilde{\Delta}_{\rm t1}(k,p)
c^{\dagger}_{1,k\downarrow}c^{\dagger}_{1,p\downarrow}+{\rm h.c.}].
\label{ham02}
\end{eqnarray}
Here $\tilde{\Delta}_{\rm s}(k,p)=\Delta_{\rm s}\sqrt{kp}[u_1(k,p)k+u_0(k,p)p]$.
The application of the unitary transformation (\ref{ut1}) and (\ref{ut2}) 
to the pairing terms of (\ref{ham02}) gives,
\begin{eqnarray}
&&\int dkdp[\tilde{\Delta}_{+}'(k,p)a^{\dagger}_{k+}a^{\dagger}_{p+}+
\tilde{\Delta}_{-}'(k,p)a^{\dagger}_{k-}a^{\dagger}_{p-}+{\rm h.c.}] \nonumber \\
&&-i\int dkdp[\tilde{i\Delta}_2'(k,p)
a^{\dagger}_{k+}a^{\dagger}_{p-}-i\tilde{\Delta}_2'(p,k)
a^{\dagger}_{k-}a^{\dagger}_{p+}+{\rm h.c.}],
\label{pairint}
\end{eqnarray}
where 
\begin{eqnarray}
\tilde{\Delta}_{\pm}'(k,p)=\tilde{\Delta}_{\pm}(k,p)
\mp\tilde{\Delta}_{\rm s}^{(s)}(k,p)s_{-}(k,p)-\tilde{\Delta}_{\rm s}^{(a)}(k,p)s_{+}(k,p),
\end{eqnarray}
\begin{eqnarray}
\tilde{\Delta}_2'(k,p)=\tilde{\Delta}_{\rm s}^{(s)}(k,p)r_{+}(k,p)+
\tilde{\Delta}_{\rm s}^{(a)}(k,p)r_{-}(k,p),
\end{eqnarray}
with $\tilde{\Delta}_{\rm s}^{(s)}(k,p)
=(\tilde{\Delta}_{\rm s}(k,p)+\tilde{\Delta}_{\rm s}(p,k))/4$ and
$\tilde{\Delta}_{\rm s}^{(a)}(k,p)
=(\tilde{\Delta}_{\rm s}(k,p)-\tilde{\Delta}_{\rm s}(p,k))/4$.
It is reasonable to postulate that in the $s$-wave gap $\tilde{\Delta}_{\rm s}$,
the symmetric part $\tilde{\Delta}_{\rm s}^{(s)}$ dominates, and 
$\tilde{\Delta}_{\rm s}^{(a)}$ can be neglected.
We write $k=k_{F\pm}+q$, $p=k_{F\pm}+q'$ with $q, q' \ll k_{F\pm}$ as before.
$\tilde{\Delta}_{\pm}'(k,p)\equiv \pm A_{\rm t}(q-q')$ is odd in $q-q'$, while
$\tilde{\Delta}_2'(k,p)\equiv A_{\rm s}(q-q')$ is even in $q-q'$. 
Fourier transforming to the coordinate space,
we introduce the quasiparticle operator,
\begin{eqnarray}
\gamma^{\dagger}=\int dx [\eta_{1+}(x)a^{\dagger}_{+}(x)+\eta_{2+}(x)a_{+}(-x)
+\eta_{1-}(x)a^{\dagger}_{-}(x)+\eta_{2-}(x)a_{-}(-x)].
\label{qo}
\end{eqnarray}
The Fourier transform of the odd-parity pairing term is
\begin{eqnarray}
\nu \int dx~ i m_{\rm t}(x)a^{\dagger}_{\nu}(x)a^{\dagger}_{\nu}(-x)+{\rm h.c.},
\qquad \nu=\pm,
\end{eqnarray}
while for the even-parity pairing term,
\begin{eqnarray}
i\int dx~ m_{\rm s}(x)a^{\dagger}_{+}(x)a^{\dagger}_{-}(-x)+{\rm h.c.}.
\end{eqnarray}
Here, $m_{\rm t}(x)$ ($m_{\rm s}(x)$) is the Fourier transform of
$A_{\rm t}(q)$ ($A_{\rm s}(q)$) 
and odd (even) in $x$.
Then, the BdG equations for 
$\Psi^T(x)=(\eta_{1+}(x),\eta_{2+}(x),\eta_{1-}(x),\eta_{2-}(x))$ are,
\begin{eqnarray}
-iv\partial_x\eta_{1+}(x)-i m_{\rm t}(x)\eta_{2+}(x)-i m_{\rm s}(x)\eta_{2-}(x)=
E \eta_{1+}(x), 
\nonumber\\
iv\partial_x\eta_{2+}(x)+i m_{\rm t}(x)\eta_{1+}(x)-i m_{\rm s}(x)\eta_{1-}(x)=
E \eta_{2+}(x), 
\nonumber\\
-iv\partial_x\eta_{1-}(x)+i m_{\rm t}(x)\eta_{2-}(x)+i m_{\rm s}(x)\eta_{2+}(x)=
E \eta_{1-}(x), 
\nonumber\\
iv\partial_x\eta_{2-}(x)-i m_{\rm t}(x)\eta_{1-}(x)+i m_{\rm s}(x)\eta_{1+}(x)=
E \eta_{2-}(x). 
\label{bdg2}
\end{eqnarray}
Here, to simplify the analysis, we have 
assumed $v_{+}\approx v_{-}\equiv v$ which is justified for
$E_F\gg \alpha$. From Eq. (\ref{bdg2}),
we find two sets of solutions with zero energy eigen value $E=0$ 
up to normalization factors,
\begin{eqnarray}
\Psi_1^{T}(x)=(C_{+}(x),C_{+}(x),-C_{-}(x),C_{-}(x)),
\label{solm1}
\end{eqnarray}
\begin{eqnarray}
\Psi_2^{T}(x)=(-iC_{-}(x),-iC_{-}(x),iC_{+}(x),-iC_{+}(x)),
\label{solm2}
\end{eqnarray}
with
\begin{eqnarray}
C_{\pm}(x)=e^{-\frac{1}{v}\int^x_{x_{+}}dy[m_{\rm t}(y)+m_{\rm s}(y)]}
\pm e^{-\frac{1}{v}\int^x_{x_{-}}dy[m_{\rm t}(y)-m_{\rm s}(y)]},
\end{eqnarray}
and $x_{\pm}$ the solution of $m_t(x_{\pm})\pm m_s(x_{\pm})=0$.
We can easily verify that $x_{-}=-x_{+}$ and $C_{+}(x)$ ($C_{-}(x)$)
is an even (odd) function of $x$.
The above solutions (\ref{solm1}) and (\ref{solm2})
are normalizable only when $m_{\rm t}(x)\pm m_{\rm s}(x)>0$ for $x>x_{\pm}$ 
and $m_{\rm t}(x)\pm m_{\rm s}(x)<0$ for $x<x_{\pm}$.
Therefore as long as the gap for the spin-triplet pairs is larger than
that for the spin-singlet pairs, the zero energy modes exist,
which is consistent with the recent result obtained by Lu and Yip \cite{yip}. 
It is noted that the quasiparticle operator (\ref{qo}) for the solutions
(\ref{solm1}) and (\ref{solm2}) 
satisfies $\gamma^{\dagger}=\gamma$;
the quasiparticles corresponding to these solutions
are Majorana fermions.

The phase with the two Majorana fermion modes is classified
as the phase I or I' in Table \ref{table:topologicalnumber}.

\subsection{Non-Abelian statistics of vortices}
\label{subsec:nonabelian}

The non-Abelian statistics of vortices is realized when there is
only one Majorana mode in a vortex core \cite{read,ivanov,Sato03}.
Thus, it is necessary to eliminate one of two Majorana fermion modes
found in Sec. \ref{subsec:index}.
For this purpose, 
we consider the case that the Fermi level crosses the $\Gamma$ point
in the Brillouin zone; i.e. $\varepsilon_{{\bm k}={\bm 0}}-\mu=0$ ($\mu=-4t$).
Then, for $H_z\neq 0$, a gap $\sim \mu_{\rm B}H_z$ opens 
in the vicinity of the $\Gamma$
point at the Fermi level~\cite{fuji1}.

In the case of purely $p$-wave pairing,
this implies that $v_{-}$ in (\ref{bdg}) vanishes, and instead
the mass term $2\sigma_z|\mu_{\rm B}H_z|\eta_{-}(x)$ is added.
In this case, there is no zero energy mode for the $\nu=-$ band,
and there is only one zero energy Majorana mode for the quasiparticles with
the Fermi momentum $k_{F+}$($\neq 0$) in the $\nu=+$ band, which
ensures the non-Abelian statistics of vortices~\cite{fuji1}.
This state is topologically equivalent to the phase II in Table
\ref{table:topologicalnumber},
and also to spinless $p+ip$ superconductivity.
However, the realization of this state in NCS is more
feasible than that of spinless $p+ip$ superconductivity,
because, for spinless $p+ip$ state, the strong magnetic field associated with
full spin polarization leads to the fatal orbital depairing effect
on superconductivity, while, for the Rashba NCS with $\mu=-4t$,
a weak magnetic field between $H_{\rm c1}$ and $H_{\rm c2}$
applied parallel to the $z$-axis is sufficient to
eliminate one of two Majorana modes.

In a similar manner,
in the case of $s+p$-wave pairing with $|\Delta_{\rm t}|>|\Delta_{\rm s}|$,
for $\mu=-4t$, there remains only one zero energy mode:
The BdG equations for the quasiparticle operator
(\ref{qo}) with $E=0$ becomes
\begin{eqnarray}
-iv\partial_x\eta_{1+}(x)-i m_{\rm t}(x)\eta_{2+}(x)
-i m_{\rm s}(x)\eta_{2-}(x)=0,
\nonumber\\
iv\partial_x\eta_{2+}(x)+i m_{\rm t}(x)\eta_{1+}(x)
-i m_{\rm s}(x)\eta_{1-}(x)=0,
\nonumber\\
|2\mu_{\rm B}H_z|\eta_{1-}(x)+i m_{\rm t}(x)\eta_{2-}(x)
+i m_{\rm s}(x)\eta_{2+}(x)=0,
\nonumber\\
-|2\mu_{\rm B}H_z|\eta_{2-}(x)-i m_{\rm t}(x)\eta_{1-}(x)
+i m_{\rm s}(x)\eta_{1+}(x)=0.
\label{eq:v_BdG}
\end{eqnarray}
When $m_{\rm t}(x)$ satisfies $m_{\rm t}(x)={\rm sgn}(x)|m_{\rm t}(x)|$,
(\ref{eq:v_BdG}) has only one normalizable solution,
\begin{eqnarray}
\eta_{1+}(x)=C(x),
\quad
\eta_{2+}=C(x),
\quad
\eta_{1-}(x)=
\frac{m_{\rm s}(x)(m_{\rm t}(x)-2i|\mu_{\rm B}H_z|)}
{4(\mu_{\rm B}H_z)^2+m_{\rm t}^2(x)},
\quad
\eta_{2-}(x)=
-\frac{m_{\rm s}(x)(m_{\rm t}(x)-2i|\mu_{\rm B}H_z|)}
{4(\mu_{\rm B}H_z)^2+m_{\rm t}^2(x)},
\end{eqnarray}
with
\begin{eqnarray}
C(x)=e^{-\frac{1}{v}\int_0^x\frac{m_{\rm t}(y)
(4(\mu_{\rm B}H_z)^2+m_{\rm t}^2(y)-m_{\rm s}^2(y))
-2im_{\rm s}^2(y)|\mu_{\rm B}H_z|}
{4(\mu_{\rm B}H_z)^2+m_{\rm t}(y)}dy}. 
\end{eqnarray}
For this solution, the quasiparticle operator is
\begin{eqnarray}
\gamma^{\dagger}=\int dx C(x)
\left[a_{+}^{\dagger}(x)+a_{+}(-x)
+\frac{m_{\rm s}(x)(m_{\rm t}(x)-2i|\mu_{\rm B}H_z|)}
{4(\mu_{\rm B}H_z)^2+m_{\rm t}^2(x)}(a_{-}^{\dagger}(x)-a_{-}(-x))\right],
\end{eqnarray}
which satisfies $\gamma^{\dagger}=\gamma$.
Thus, there is only one Majorana zero energy mode in a vortex core.
Under this situation, the non-Abelian statistics of vortices is possible.

\subsection{Majorana condition}

The Majorana condition of zero energy vortex core states is crucial to
the non-Abelian statistics of the vortices, so it is better to argue it
without any approximation.
In this section, we present a general argument on the Majorana condition
of vortex zero modes for the 2D NCS.

Let us start with (\ref{ham02}) in the Nambu representation,  
\begin{eqnarray}
{\cal H}^{(01)}=\frac{1}{2}\int \frac{dkdp}{(2\pi)^2}
\left(c_{0,k\uparrow}^{\dagger},c_{1,k\downarrow}^{\dagger},
c_{0,k\uparrow},c_{1,k\downarrow}\right){\cal H}(k,p)
\left(
\begin{array}{c}
c_{0,p\uparrow} \\
c_{1,p\downarrow}\\
c_{0,p\uparrow}^{\dagger}\\
c_{1,p\downarrow}^{\dagger}
\end{array}
\right), 
\end{eqnarray}
where ${\cal H}(k,p)$ is given by
\begin{eqnarray}
{\cal H}(k,p)=
\left(
\begin{array}{cccc}
\xi_{k\uparrow}\delta_{k,p} &i\alpha k \delta_{k,p} 
& h^{01}_{{\rm t}\uparrow}(k,p)& h^{01}_{\rm s}(k,p)\\
-i\alpha k \delta_{k,p} & \xi_{k\downarrow}\delta_{k,p}
& -h^{01}_{\rm s}(p,k) & h^{01}_{{\rm t}\downarrow}(k,p)\\
h^{01*}_{{\rm t}\uparrow}(p,k) & -h^{01*}_{\rm s}(k,p) 
& -\xi_{k\uparrow}\delta_{k,p} & i\alpha k \delta_{k,p}\\
h^{01*}_{\rm s}(p,k) & h^{01*}_{{\rm t}\downarrow}(p,k)&
-i\alpha k \delta_{k,p} & -\xi_{k\downarrow}\delta_{k,p}
\end{array}
\right)
\label{eq:BdGzeromode}
\end{eqnarray}
with
\begin{eqnarray}
h^{01}_{\rm s}(k,p)=-i(2\pi)^2\tilde{\Delta}_{\rm s}(k,p),\quad
h^{01}_{{\rm t}\uparrow}(k,p)=2(2\pi)^2\tilde{\Delta}_{\rm t0}(k,p)
\quad
h^{01}_{{\rm t}\downarrow}(k,p)=-2(2\pi)^2\tilde{\Delta}_{\rm t1}(k,p).
\nonumber\\
\end{eqnarray}
Using the following relation,
\begin{eqnarray}
\left(
\begin{array}{c}
c_{0,p\uparrow} \\
c_{1,p\downarrow}\\
c_{0,p\uparrow}^{\dagger}\\
c_{1,p\downarrow}^{\dagger}
\end{array}
\right)
=
\Gamma
\left(
\begin{array}{c}
c_{0,p\uparrow} \\
c_{1,p\downarrow}\\
c_{0,p\uparrow}^{\dagger}\\
c_{1,p\downarrow}^{\dagger}
\end{array}
\right)^{*},
\quad
\Gamma=\left(
\begin{array}{cc}
0 &{\bm 1}_{2\times 2} \\
{\bm 1}_{2\times 2} &0
\end{array}
\right), 
\end{eqnarray}
one can show that ${\cal H}(k,p)$ has the so-called particle hole symmetry,
\begin{eqnarray}
\Gamma {\cal H}(k,p)\Gamma =-{\cal H}^{*}(k,p),
\label{eq:particle-hole}
\end{eqnarray}
which also can be checked directly from (\ref{eq:BdGzeromode}).

In the momentum space, the BdG equation is given by,
\begin{eqnarray}
\int dp {\cal H}(k,p){\bm u}(p)=E{\bm u}(k),
\end{eqnarray}
where ${\bm u}(p)$ is a four component vector 
${\bm u}(p)=(\alpha(p),\beta(p),\gamma(p),\delta(p))^{\rm t}$.
When ${\bm u}(p)$ is a solution of the BdG equation with the energy $E$,
then, using (\ref{eq:particle-hole}), we can show that $\Gamma{\bm u}^{*}(p)$ 
is a solution with the energy $-E$, 
\begin{eqnarray}
\int dp {\cal H}(k,p)\Gamma {\bm u}^{*}(p)=-E \Gamma {\bm u}^{*}(k). 
\end{eqnarray}
Therefore, if ${\bm u}_0(p)$ is a zero mode of the BdG equation, then 
$\Gamma{\bm u}_0^{*}(p)$ is also a zero mode.
This means that if the BdG equation has only one
independent zero mode ${\bm u}_0(p)$, then ${\bm u}_0(p)$ and 
$\Gamma {\bm u}_0(p)$ should not be independent of each other.
In general, if there are an odd number of independent zero modes,
then at least for one solution, ${\bm u}_0(p)$ and 
$\Gamma{\bm u}_0^{*}(p)$ are not independent.

For simplicity, suppose that the BdG equation has a single zero mode
${\bm u}_0(p)$.
As mentioned above, ${\bm u}_0(p)$ and $\Gamma {\bm u}_0^{*}(p)$ are not
independent, so we have 
\begin{eqnarray}
{\bm u}_0(p)=\Gamma {\bm u}_0^{*}(p), 
\end{eqnarray}
by choosing a suitable phase of ${\bm u}_0(p)$.
In terms of the components of ${\bm u}_0(p)$, this becomes
\begin{eqnarray}
{\bm u}_0(p)\equiv
\left(
\begin{array}{c}
\alpha_0(p) \\
\beta_0(p)\\
\chi_0(p)\\
\eta_0(p)
\end{array}
\right)
=
\left(
\begin{array}{c}
\chi_0^{*}(p) \\
\eta_0^{*}(p)\\
\alpha_0^{*}(p)\\
\beta_0^{*}(p)
\end{array}
\right). 
\end{eqnarray}
To obtain the annihilation operator $\gamma$ for the zero
mode, we perform the mode expansion for $(c_{0p\uparrow}, c_{1p\downarrow},
c_{0p\uparrow}^{\dagger}, c_{1p\downarrow}^{\dagger})$, 
\begin{eqnarray}
\left(
\begin{array}{c}
c_{0,p\uparrow} \\
c_{1,p\downarrow}\\
c_{0,p\uparrow}^{\dagger}\\
c_{1,p\downarrow}^{\dagger}
\end{array}
\right)
=\gamma
\left(
\begin{array}{c}
\alpha_0(p) \\
\beta_0(p)\\
\chi_0(p)\\
\eta_0(p)
\end{array}
\right)+\cdots,
\end{eqnarray}
where we have omitted the terms including non-zero modes. 
In this equation, we have two dependent relations,
\begin{eqnarray}
c_{0,p\uparrow}=\gamma \alpha_0(p)+\cdots,
\quad
c_{0,p\uparrow}^{\dagger}=\gamma \chi_0(p)+\cdots, 
\nonumber\\
c_{1,p\downarrow}=\gamma \beta_0(p)+\cdots,
\quad
c_{1,p\downarrow}^{\dagger}=\gamma \eta_0(p)+\cdots.
\end{eqnarray}
Thus $\gamma$ must satisfy the Majorana condition,
\begin{eqnarray}
\gamma=\gamma^{\dagger}. 
\end{eqnarray}
From the normalization condition for the zero mode, 
$\gamma$ can be written as
\begin{eqnarray}
\gamma=\int dp(\alpha_0^*(p),\beta_0^*(p),\chi_0^*(p),\eta_0^*(p))
\left(
\begin{array}{c}
c_{0,p\uparrow} \\
c_{1,p\downarrow}\\
c_{0,p\uparrow}^{\dagger}\\
c_{1,p\downarrow}^{\dagger}
\end{array}
\right).
\label{eq:zeromode2}
\end{eqnarray}
Thus the commutation relation of $\eta_0$ can be calculated as
\begin{eqnarray}
\{\gamma,\gamma^{\dagger}\} 
&=&\int dk dp \{\alpha_0^{*}(k)c_{0,k\uparrow}+\beta_0^{*}(k)c_{1,k\downarrow}
+\chi_0^{*}(k)c_{0,k\uparrow}^{\dagger}
+\eta_0^{*}(k)c_{1,k\downarrow}^{\dagger},
\nonumber\\
&&\hspace{5ex}
\alpha_0(p)c_{0,p\uparrow}^{\dagger}+\beta_0(p)c_{1,p\downarrow}^{\dagger}
+\chi_0(p)c_{0,p\uparrow}+\eta_0(p)c_{1,p\downarrow}
\}
\nonumber\\
&=&\int dk dp \left[
\alpha_0^{*}(k)\alpha_0(p)
+\beta_0^{*}(k)\beta_0(p)
+\chi_0^{*}(k)\chi_0(p)
+\eta_0^{*}(k)\eta_0(p)
\right]\delta_{k,p}
\nonumber\\
&=&1
\end{eqnarray}
In a similar manner, we can show that $\gamma$ anti-commutes with
annihilation and creation operators for non-zero modes.

In conclusion, the existence of only one zero energy mode in a vortex core is
the necessary and sufficient condition for the existence of
a single Majorana mode which leads to the non-Abelian statistics.

\section{Summary}
\label{sec:summary}

We have explored topological phases of NCS characterized by 
the existence of gapless
edge states and Majorana fermion modes in vortex cores, mainly focusing
on the 2D Rashba superconductors with the admixture of $s$-wave pairing and
$p$-wave pairing.
It has been clarified that when the $p$-wave gap is larger than
the $s$-wave gap, topological states are realized.
We have completed 
the topological classification of these states by examining
the TKNN number, the ${\bm Z}_2$ index, and the winding number associated
with specific symmetry points in the Brillouin zone.
It has been also found that for the Rashba
superconductors, the gapless edge states that ensure
the quantum spin Hall effect protected from
disorder are stable against
a weak magnetic field applied perpendicular 
to the propagating direction of the edge states,
despite broken time-reversal symmetry which flaws
the ${\bm Z}_2$ characterization of the topological phase.
The stability of the edge states is guaranteed by
a specific accidental symmetry of the Rashba model.
We have also proposed a simple scheme for the realization of
the non-Abelian statistics of vortices in topological phases of NCS
under an applied magnetic field.

Experimental verification of these findings are particularly of interest.
The topologically-protected gapless edge states play important roles for
transport properties.
In the superconducting state, current flows carried by the edge states
can be detected by the measurement of a spin Hall current
or thermal transport measurements at sufficiently low temperatures 
where bulk quasiparticles are well suppressed.
Also, the accidental topological phase in the case with a magnetic field
can be detected by observing the dependence of the transport current
on the direction of the magnetic field, or the splitting of the zero-energy
bias peak of a tunneling conductance due to the tilt of the magnetic field.
More elaborate argument on these experimental implications
will be addressed in the near future.

The experimental realization of the non-Abelian statistics of vortices
is most challenging,
though the scheme proposed in this paper is, in principle, feasible.
The $s+p$-wave NCS with the Fermi level tuned to be $\mu=-4t$ discussed in 
Sec. \ref{subsec:nonabelian} need not to be bulk systems.
A proximity-induced superconductor realized 
in the vicinity of the interface between
a $p$-wave superconductor and a semiconductor with the Rashba SO interaction
may be a good candidate for the realization of this phenomenon.
For the experimental detection of the non-Abelian statistics,
the two-point-contact interferometer 
considered in Refs.\cite{fradkin,fuji1,bonderson,stern} may be employed.

\begin{acknowledgments}
The authors are grateful to S. K. Yip for invaluable discussions.
They also thank the organizers of the symposium, ``Topological Aspects
of Solid State Physics'', at YITP, Kyoto, where this work has been
 started.
This work is partly supported by the Grant-in-Aids for
the Global COE Program
``The Next Generation of Physics, Spun from Universality and Emergence''
and for Scientific Research 
(Grant No.18540347, Grant No.19014009, Grant No.19052003)
from MEXT of Japan.

\end{acknowledgments}

\bibliography{edge-vortex-NCS}% Produces the bibliography via BibTeX.

\end{document}